\documentclass[pra,amsmath,amssymb,twocolumn,longbibliography]{revtex4-1}
\usepackage[dvips,letterpaper,margin=0.5in,bottom=0.5in, top=0.5in]{geometry} 
\usepackage[english]{babel}
\usepackage{epstopdf}
\usepackage{float}
\usepackage{enumerate}

\usepackage{amsmath} 
\usepackage{dsfont}
\usepackage{cancel}
\usepackage{amsthm} 
\usepackage{amssymb}	
\usepackage{amsmath}
\usepackage{graphicx} 
\usepackage{empheq}
\usepackage[absolute,overlay]{textpos}
\usepackage{xcolor}
\usepackage{hyperref}
\usepackage{tikz}
\usepackage{relsize}

\begin{document}
\title{Stability of quasicrystalline ultracold fermions to dipolar interactions}
\date{\today}
\author{Paolo Molignini}
\affiliation{Department of Physics, Stockholm University, AlbaNova University Center, 106 91 Stockholm, Sweden}

\begin{abstract}
Quasiperiodic potentials can be used to interpolate between localization and delocalization in one dimension. 
With the rise of optical platforms engineering dipolar interactions, a key question is the stability of quasicrystalline phases under these long-range interactions.
In this work, we study repulsive ultracold dipolar fermions in a quasiperiodic optical lattice to characterize the behavior of interacting quasicrystals.
We simulate the full time evolution of the typical experimental protocols used to probe quasicrystalline order and localization properties.
We extract experimentally measurable dynamical observables and correlation functions to characterize the three phases observed in the noninteracting setting: localized, intermediate, and extended.
We then study the stability of such phases to repulsive dipolar interactions.
We find that dipolar interactions can completely alter the shape of the phase diagram by stabilizing the intermediate phase, mostly at the expense of the extended phase.
Moreover, in the strongly interacting regime, a resonance-like behavior characterized by density oscillations appears.
Remarkably, strong dipolar repulsions can also localize particles even in the absence of quasiperiodicity if the primary lattice is sufficiently deep.
Our work shows that dipolar interactions in a quasiperiodic potential can give rise to a complex, tuneable coexistence of localized and extended quantum states.
\end{abstract}
\maketitle


\emph{Introduction} ---
Since their revolutionary discovery in aluminum-manganese alloys~\cite{Shechtman:1984}, quasicrystals have been the subject of intense theoretical and experimental research due to their many exotic properties.
Quasicrystals exhibit a regular structure despite lacking a canonical periodicity~\cite{Rabson:1991,Goldman:1993}.
While they lack translational symmetry, they can completely tesselate a two-dimensional surface in an aperiodic fashion because they do possess unexpected rotational symmetries (five-fold, eight-fold, etc.).
Quasicrystals can be mathematically understood in terms of higher-dimensional periodic structures projected onto lower-dimensional slices, and have thus ushered in new paradigms for crystallography~\cite{Mackay:1995,Yamamoto:1996,Steurer:2008,Steurer:2009,Wolny:2018}, band theory in solids~\cite{Poon:1992,Kelton:1993,Rotenberg:2004,Grimm:2003}, and hidden dimensions~\cite{Jagannathan:2021}.
Moreover, quasiperiodic structures can emulate on-site disorder, which makes them ideal platforms to study many-body localization~\cite{Iyer:2013,Schreiber:2015,Setiawan:2017,Khemani:2017,Zhang:2018,Mace:2019,Varma:2019,Abanin:2019,Agrawal:2022} and its interplay with nonergodicity~\cite{Li:2015,Li:2016,Hsu:2018,Ghosh:2020,Roy:2021,Roy:2023}, integrability~\cite{He:2013,Znidaric:2018,Singh:2021,Znidaric:2021,Thomson:2023}, and fractality~\cite{Thiem:2011, Kalugin:2014, Mace:2016, Mace:2017, Rai:2019, Rai:2020, Jagannathan:2021, Dai:2023}.

In recent years, impressive advances in trapping and controlling ultracold atoms have made them an excellent platform where to realize and study quasiperiodicity with extreme precision~\cite{Lye:2007, Fallani:2007, Singh:2015, Schreiber:2015, Mace:2016-2, Bordia:2017, Lueschen:2017, Lueschen:2017-2, Lueschen:2018, Corcovilos:2019, Viebahn:2019, Johnstone:2019, Sbroscia:2020, Gautier:2021,Nakajima:2021,Yu:2023}.
In one dimension (1D), a quasiperiodic potential can be generated by superimposing a primary optical lattice with a detuning lattice at an incommensurate frequency.
Upon loading particles in such a quasiperiodic potential, it is possible to quantum simulate and study the dynamics of 1D quasicrystals with high accuracy.
A particularly interesting feature exhibited by these systems is the coexistence of localized and extended states in an intermediate phase separating the fully localized to the fully extended phase~\cite{Lueschen:2018}.
In the intermediate phase, the different states are separated by a critical energy level termed single-particle mobility edge (SPME)~\cite{Abrahams:1979, Lee:1985, Boers:2007, Ganeshan:2015, Li:2017, Lueschen:2018}.
This coexistence manifests itself due to long-range tunneling terms that appear in the continuum description of the 1D ultracold quantum simulator.
In fact, the SPME is absent in the deep lattice limit, where an effective Aubry-Andr\'{e} model with self-duality is obtained~\cite{Aubry:1980, Fallani:2007, Roati:2008, Modugno:2009}.
Furthermore, a rigorous SPME does not appear in 1D systems with Anderson-type (uncorrelated) disorder~\cite{Lueschen:2018}.

While the properties of the SPME and intermediate phase have been widely explored in noninteracting systems, less attention has been given to potential interplays with interactions, in particular long-range ones that can arise in the new generation of dipolar quantum simulators~\cite{Micheli:2006, Lahaye:2009, Baranov:2012,Gross:2017}.
In fact, nowadays numerous atomic and molecular species exhibiting dipole-dipole interactions (DDI) are available in ultracold labs, such as dysprosium $^{161}$Dy~\cite{Lu:2012}, erbium $^{167}$Er~\cite{Aikawa:2014, Baier:2018},  chromium$^{53}$Cr~\cite{Naylor:2015}, potassium-rubidium $^{40}$K$^{87}$Rb~\cite{Ni:2008}, sodium-lithium $^{23}$Na$^6$Li~\cite{Rvachov:2017}, and sodium-potassium $^{23}$Na$^{40}$K~\cite{Park:2015, Duda:2023}.
DDI, beside better incorporating the long-range nature of interactions in solid-state quasicrystalline materials, could potentially herald many new interesting phases of matter when interfaced with quasicrystalline long-range order~\cite{Bai:2015,Pandey:2020, Vu:2022}.

In this work, we investigate the role of DDI in 1D ultracold fermions loaded in a quasiperiodic potential.
We simulate the exact experimental protocol used in ultracold atomic setups~\cite{Schreiber:2015, Lueschen:2018, Kohlert:2019, An:2021}, whereby the fermions are first loaded in a superlattice of twice the periodicity of the primary optical lattice, and then are let to time evolve into the quasiperiodic structure quenched at time zero.
By measuring the imbalance between odd and even sites and the expansion of the particle density, it is possible to classify the state of the system as being in a localized, extended, or intermediate phase.
We study the stability of each phase with respect to DDI by determining phase diagrams and by calculating observables such as pair-correlation functions.

We find that DDI tend to stabilize the intermediate phase, mostly to the detriment of the extended phase.
Examining the correlation functions, this stabilization results from a decreased correlation between far away particles that can be attributed to interference between the long-range tails of the DDI. 
At strong DDI, the same mechanism not only strengthens the imbalance in the localized phase, but also induce a new kind of localization in regimes where the noninteracting system would only host extended states.
Surprisingly, we find that this localization persists also when the detuning lattice is zero, i.e. for a fully periodic optical lattice.
By combining the effect of quasiperiodicity and DDI-induced bound clusters, we bring forth an explanation for the pathway from disorder-free localization, to delocalization, to eventual re-localization.
Our study illustrates the potential for long-range interacting systems to induce new types of localization phenomena in conjunction with quasicrystalline structures.

\emph{Model} ---
We consider the dynamics of a system of $N$ ultracold fermions placed in a one-dimensional quasiperiodic optical lattice.
The fermions are described by creation/annihilation field operators $\hat{\Psi}(x)^{(\dagger)}$ obeying canonical anticommutation relations and their single particle Hamiltonian is given by
\begin{align}
\mathcal{H} &= \int \mathrm{d}x \: \hat{\Psi}^{\dagger}(x) \left[ -\frac{\hbar^2}{2m} \frac{\partial^2}{\partial x^2}  + V(x) \right] \hat{\Psi}(x).
\end{align}
The quasiperiodic lattice $V(x)$ is realized by interfering a detuning laser of amplitude $V_d$, wave vector $k_d$, and phase $\phi$, with a primary laser of amplitude $V_p$ and wave vector $k_p$,
\begin{equation}
V(x) = \frac{V_p}{2 E_r} \cos (2 k_p x) + \frac{V_d}{2 E_r} \cos(2 k_d x + \phi),
\end{equation}
with $E_r$ the recoil energy of the primary lattice (unit of energy). 
For the potential parameters we choose values compatible with experimental realizations of this system~\cite{Lueschen:2018,supmat}.
We also add hard wall boundaries~\cite{Gaunt:2013,Mazurenko:2017,Gall:2021,Navon:2021} to obtain a finite system containing 64 sites in the primary lattice.
While this spatial extent is smaller than in typical experiments, it is more than enough to capture the hallmarks of each phase.

The fermions interact with one another with the interaction Hamiltonian
\begin{align}
\mathcal{H}_{\mathrm{int}} &=  \frac{1}{2} \int \mathrm{d}x \: \int \mathrm{d}x' \: \hat{\Psi}^{\dagger}(x) \hat{\Psi}^{\dagger}(x')  W(x,x') \hat{\Psi}(x') \hat{\Psi}(x),
\end{align}
where 
\begin{equation}
W(x,x') = \frac{W \bar{L}^3}{E_R( |x-x'|^3 + \alpha)},
\end{equation}
is a dipolar repulsion with interaction strength $W$ and regularization factor $\alpha=0.01$ (phenomenologically accounting for Pauli pressure), and $\bar{L}$ is the distance between two neighboring minima in the primary lattice
(unit of length).

\emph{Experimental protocol and observables} --- To probe the properties of the system, we follow the  protocol employed in previous experiments~\cite{Schreiber:2015, Lueschen:2018}.
We first build a superlattice with half the primary lattice wave vector and load particles into its center.
At time zero, we then quench the primary and detuning lattices to their respective values and let the system time evolve.
We monitor the real time evolution and extract dynamical observables from the many-body wave function $\left| \Psi(t) \right>$ of the system.
To characterize the localization properties of the system, we calculate two dynamical quantities: \emph{imbalance} $\mathcal{I}(t)$ and \emph{edge density} $\mathcal{D}(t)$.
The imbalance is defined as the instantaneous normalized density difference between the odd and even sites in the superlattice:
\begin{equation}
\mathcal{I}(t) \equiv \frac{\mathcal{N}_e(t) - \mathcal{N}_o(t)}{\mathcal{N}_e(t) + \mathcal{N}_o(t)},
\end{equation}
where $\mathcal{N}_{e/o}(t) \equiv \int_{e/o} \mathrm{d}x \: \rho(x,t)$, $\rho(x,t) = \left< \Psi(t) \right| \hat{\Psi}^{\dagger}(x) \hat{\Psi}(x) \left| \Psi(t) \right>$, and $e$ resp. $o$ indicate the spatial points belonging to the even or odd superlattice sites.
The edge density is defined as the ratio of particles outside the central range populated at time zero:
\begin{equation}
\mathcal{D}(t) \equiv 1 - \mathcal{N}_c/N
\end{equation}
with $\mathcal{N}_c = \int_{c} \mathrm{d} x \: \rho(x,t)$ and $c$ referring to the initially populated central region (e.g. for $N=4$, $c=[-4.5,3.5]$).
These two observables in the long-time limit help us to determine the many-body phase of the system.
A finite imbalance $\mathcal{I}$ indicates the persistence of the initial charge density wave pattern imposed by the superlattice and is a proxy for the presence of localized states.
A finite (and growing) value of the edge density $\mathcal{D}$ reveals instead the presence of extended states.
Therefore, in the localized phase $\mathcal{I} > 0$ and $\mathcal{D} = 0$, while in the extended phase $\mathcal{I} = 0$ and $\mathcal{D} > 0$.
Localized and extended states can also coexist at different energies in an intermediate phase, where both values of $\mathcal{I}$ and $\mathcal{D}$ are nonzero and a SPME is present.
While the SPME should occur at arbitrarily small nonzero $\mathcal{I}$ and $\mathcal{D}$, numerical simulations introduce small imprecisions.
Thus small but finite thresholds $t_{\mathcal{I}}$ and $t_{\mathcal{D}}$ for the imbalance and the edge density have to be introduced.
Throughout our analysis, we will empirically set them to be $t_{\mathcal{I}}=0.1$ and $t_{\mathcal{D}}=0.03$~\footnote{We found empirically that these threshold values minimize numerical deviations from the experimentally observed phase diagram.}.

Besides exploring how interactions impact the phase diagram and the boundaries between different phases, we also probe the correlations within each phase.
For simplicity, we focus here on the behavior of the diagonal of the reduced two-body density matrix (2-RDM), defined as
\begin{equation}
\rho^{(2)}(x,x'; t) = \left<\Psi(t) \right| \hat{\Psi}^{\dagger}(x) \hat{\Psi}^{\dagger}(x') \hat{\Psi}(x') \hat{\Psi}(x) \left| \Psi(t) \right>,
\end{equation}
where $\left| \Psi(t) \right>$ is the time-evolved many-body state of the system.
The diagonal of the 2-RDM is equivalent to the pair-correlation function and describes the conditional probability of finding a fermion at position $x$, when another fermion is located at $x'$, i.e. it quantifies how a fermions is surrounded by other fermions.
Thus, it encodes the entire information about pairwise interactions among particles and is a proxy for the (unnormalized) pairwise correlations of the system.

\emph{Methods} --- To simulate the full many-body interacting system, we employ the MultiConfigurational Time-Dependent Hartree method for indistinguishable particles~\cite{Streltsov:2006, Streltsov:2007, Alon:2007, Alon:2008}, implemented in the MCTDH-X software~\cite{Lode:2016,Fasshauer:2016,Lin:2020,Lode:2020,MCTDHX}.
With it, we solve the many-body Schr\"{o}dinger equation directly for the continuum system consisting of kinetic energy, interparticle interactions, and optical lattice.
MCTDH-X has been widely used to study ground-state and dynamical properties of long-range interacting systems~\cite{Fischer:2015,Molignini:2018,Chatterjee:2018,Lin:2019,Chatterjee:2019,Chatterjee:2020,Lin:2020-2,Lin:2021,Molignini:2022,Hughes:2023,Bilinskaya:2024,Molignini:1-2024,Molignini:3-2024}.
MCTDH-X relies on a time-dependent variational optimization procedure in which the many-body wavefunction is decomposed into an adaptive basis set of $M$ time-dependent single-particle functions called \emph{orbitals}. 
Due to the Pauli exclusion principle, to correctly describe a set of $N$ fermions, we require $M \ge N+1$ orbitals~\footnote{Strictly speaking, $M=N$ would suffice but this case is often degenerate.}.
For strongly interacting systems, a larger number of orbitals might be required to capture many-body correlation effects.
In this work, we have verified that including orbitals beyond $M=N+4$ leads to a negligible population for the time scales probed by our dynamics~\cite{supmat}.

\begin{figure}
\centering
\includegraphics[width=0.92\columnwidth]{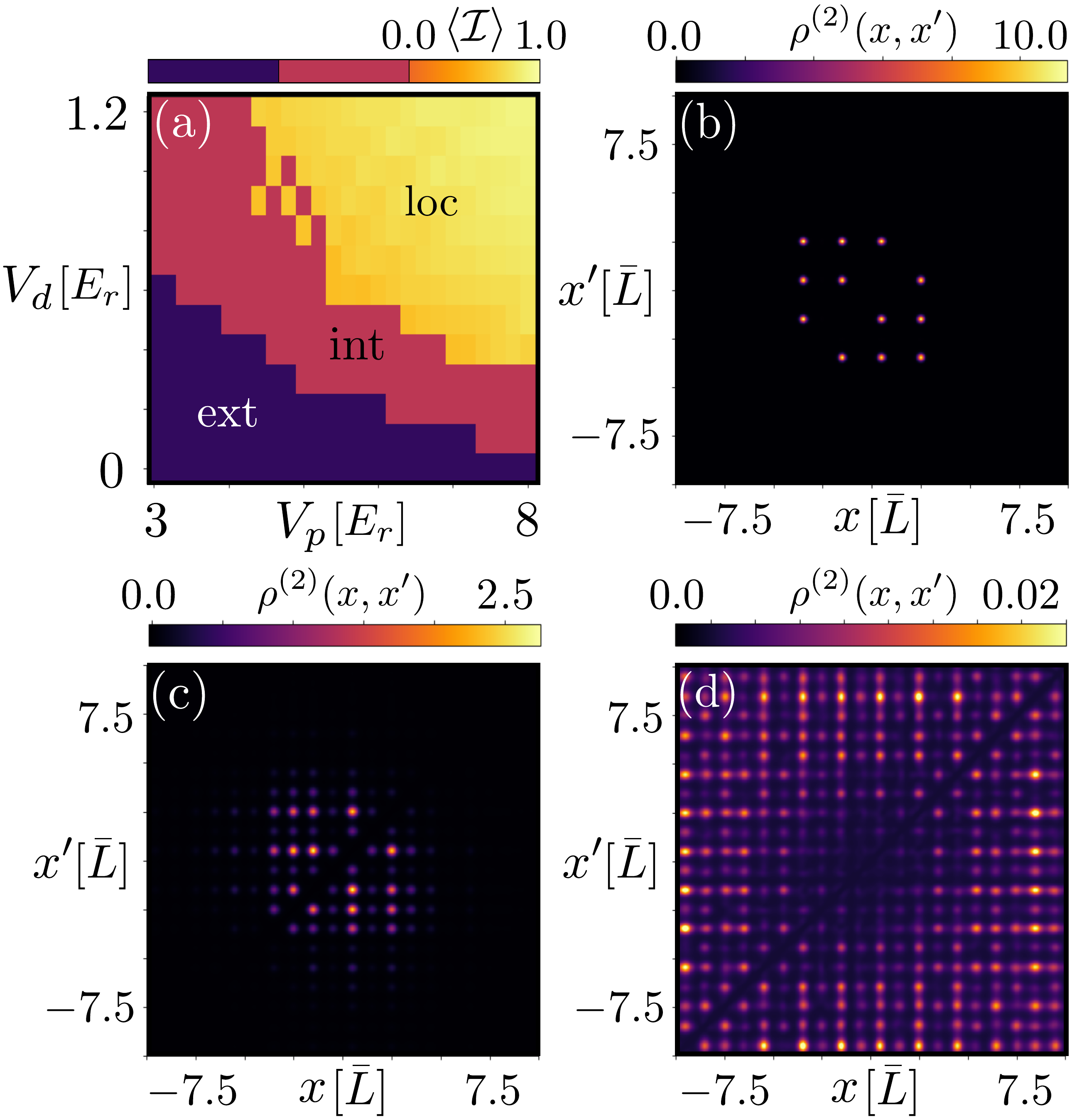}
\caption{
\textbf{Behavior of the noninteracting system}:
(a) Phase diagram for $N=4$ fermions with $M=9$ orbitals in a 1D quasiperiodic potential.
For the localized phase, the time-averaged imbalance $\left< \mathcal{I} \right>$ is plotted in a continuous color scheme.
(b)-(d) 2-RDM at $t=20 \bar{t}=3.7$ ms for
(b) localized phase ($V_p=16.0 E_r$, $V_d=2.4 E_r$), 
(c) intermediate phase ($V_p=5.0 E_r$, $V_d=0.4 E_r$), 
and (d) extended phase ($V_p=3.0 E_r$, $V_d=0 E_r$).
}
\label{fig:pd}
\end{figure}

\emph{Results for noninteracting fermions} --- We begin by describing the dynamics for noninteracting particles. 
Fig.~\ref{fig:pd}(a) shows the noninteracting phase diagrams for $N=4$ fermions with $M=9$ orbitals, constructed by calculating the imbalance and expansion observables as the depth of the primary and detuning lattices are varied.
We use parameter ranges compatible with experimental regimes, i.e. $V_p \in [3 E_r, 8 E_r]$ and $V_d \in [0, 1.2 E_r]$.
The imbalance is calculated by averaging over the last 50 time steps at the end of each time evolution, with maximal simulation times reaching $\approx 45$ ms.
The edge density is typically calculated at a target time $\tilde{t}$, well before the particles can reach the boundaries of the system.
Typical target times are  $\tilde{t} = 1.8$ ms to $\tilde{t} = 7.2$ ms~\footnote{This measurement timescale is slightly different than the experimental one, where expansion measurements are calculated at longer times. This is necessary to avoid simulating very large spatial grids that are computationally unattainable. Nevertheless, we have verified that the value of the target time $\bar{t}$ only slightly impacts the precise location of the phase boundary between the intermediate and localized phase, and that the qualitative behavior is the same for different values of $\bar{t}$.}.
From Fig.~\ref{fig:pd}(a), we can clearly distinguish the three expected phases, which have the same structure observed in prior studies~\cite{Li:2017, Lueschen:2018}.
At low values of $V_d$, the system is always in the extended phase (purple region).
Increasing the value of $V_d$ at low to moderate values of $V_p$ pushes the system into an intermediate region where extended and localized states coexist (magenta region).
Further increasing $V_d$ or $V_p$ eventually localizes all the states (yellow region).

The three phases of the quasicrystal system can also be distinguished by the dynamics of their many-body correlations encoded in the 2-RDM, shown for selected times in Fig.~\ref{fig:pd}(b)-(d) (see supplementary materials for more time snapshots~\cite{supmat}).
The 2-RDM in the localized phase [Fig.~\ref{fig:pd}(b)] does not change at all with time and merely shows the expected exchange hole at $x=x'$ with strong many-body correlation between the sites of the initial superlattice.
On the contrary, the extended phase [Fig.~\ref{fig:pd}(d)] exhibits a rapid spread of correlations at very short times already.
Between the two extremes of localization and expansion lies the intermediate phase [Fig.~\ref{fig:pd}(c)].
The correlation pattern here reveals an almost immediate exchange correlation between neighboring sites within the populated area of the initial superlattice, and a relatively small spread to sites outside of it.
However, the intensity of the 2-RDM is not uniform: correlations between sites in the initial superlattice structure still dominate throughout the time evolution even for longer times~\cite{supmat}.
These results indicate that the intermediate phase is characterized by two types of correlations showcasing the coexistence of localized and extended states.
Localized states mediate static, short-range correlations between the initially populated sites, while extended states mediate long-range correlations spanning many sites in the quasiperiodic lattice.

\emph{Results for interacting fermions} ---
We now address how the DDI impact the various quasicrystalline phases.
The protocol studied is analogous to the one used in the noninteracting case.
For simplicity, in our simulations we turn off the interactions during the initial loading procedure.
In an experiment, dipolar fermions could be directly loaded in their starting position by employing optical tweezers or letting them relax into the superlattice with a very large depth.
We again compute imbalance $\mathcal{I}$ and edge density $\mathcal{D}$ to obtain measures of localization and expansion properties, and collect them as a function of lattice depths and interaction strength.
We also calculate densities, correlations, and energies~\cite{supmat} to better describe the behavior of each phase.
\begin{figure}[t!]
\centering
\includegraphics[width=1.0\columnwidth]{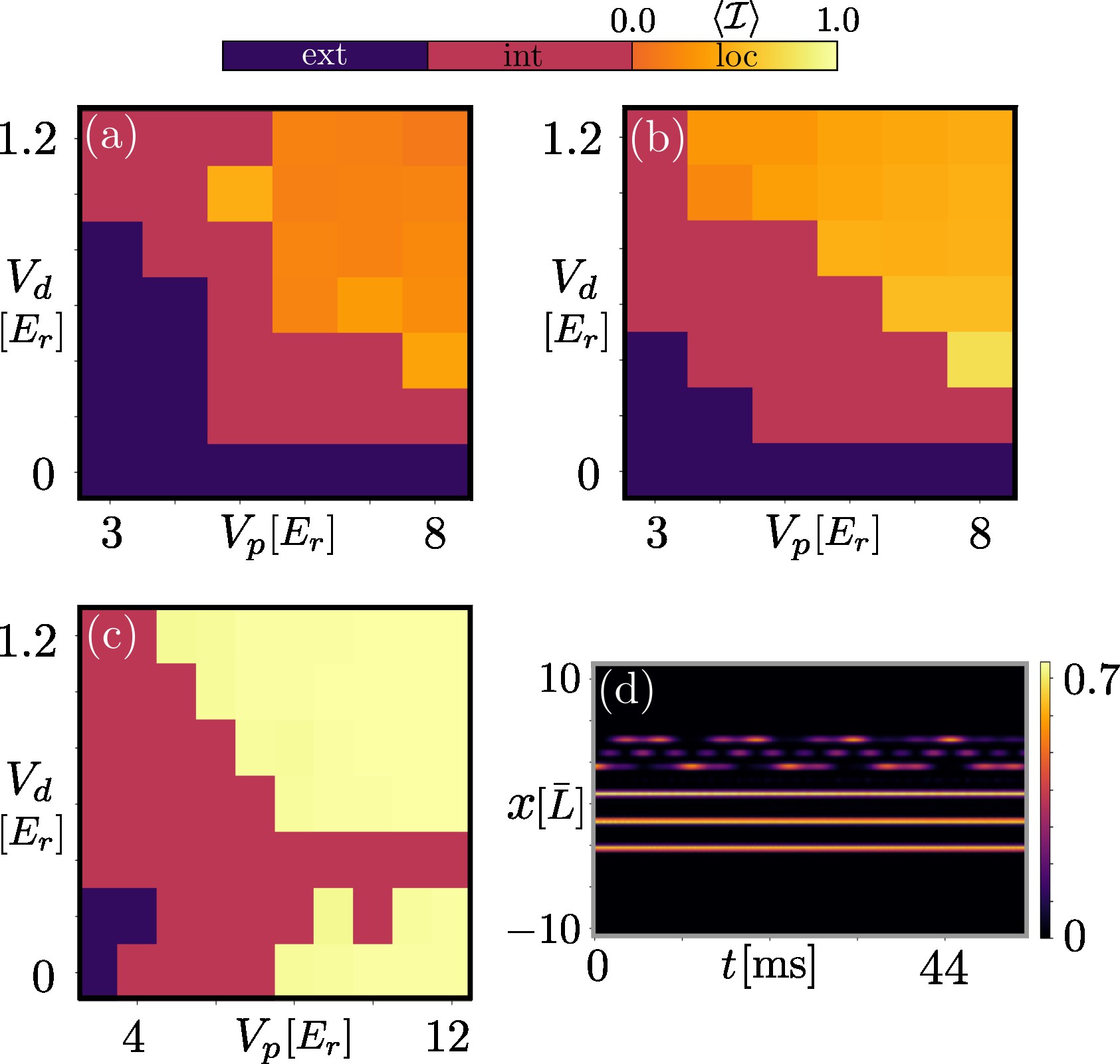}
\caption{
\textbf{Phase diagram for increasing DDI strength $W$}:
(a) $W=0.02$$E_r$,
(b) $W=0.2$$E_r$,
(c) $W=2.0$$E_r$.
For the localized phase, the time-averaged imbalance parameter $\left< \mathcal{I} \right>$ is plotted as a continuous color scheme.
Panel (d) depicts the density dynamics of the resonance lobe at $V_p= 6 E_r$, $V_d = 0.4 E_r$.
}
\label{fig:PD-p-3-pos-W}
\end{figure}

Fig.~\ref{fig:PD-p-3-pos-W} shows how the noninteracting phase diagram changes when interactions are progressively increased [panels (a) to (c)].
At small interaction values, the intermediate phase only grows slightly while the imbalance in the localized phase decreases.
As the interactions become stronger, though, the phase diagram changes drastically and in a non-monotonic fashion.
For $W=0.2 E_r$, the intermediate phase grows sizeably at the expense of most of the localized phase and part of the extended phase.
Therefore, DDI \emph{stabilize} the coexistence of extended and localized states in the intermediate phase.
At even stronger interactions, $W=2.0 E_r$, the intermediate phase dominates the probed parameter region.
Note that in Fig.~\ref{fig:PD-p-3-pos-W}(c) $V_p$ is plotted to larger values up to $12 E_r$.
The purely extended region is instead reduced to a tiny pool around $V_p=3 E_r$, $V_d=0 E_r$.

\begin{figure}[h!]
\centering
\includegraphics[width=\columnwidth]{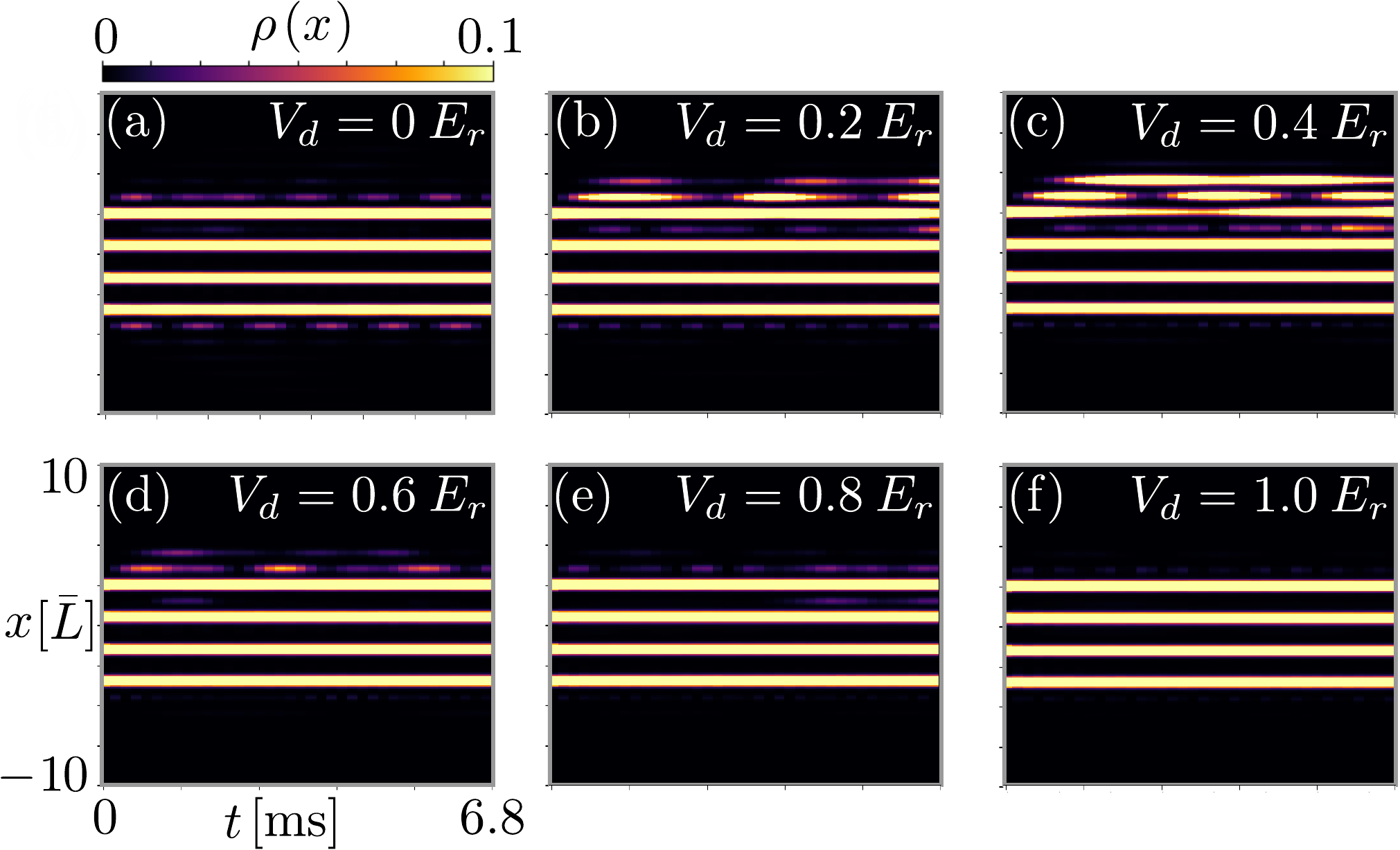}
\caption{
\textbf{Two types of localization at strong interactions.}
Density dynamics at $V_p=8 E_r$ and $W=2.0 E_r$ for increasing values of $V_d$ between 0 and $1.0 E_r$, illustrating how correlated disorder impacts the DSBS.
The scale is zoomed in to increase resolution.
}
\label{fig:dens-resonance}
\end{figure}

Most strikingly, however, a \emph{new} localized region emerges at low values of $V_d$.
Note that this region was initially part of the extended phase in the noninteracting case.
The localized phases at low and high $V_d$ are divided by a strip of intermediate phase which persists for very large values of $V_p$ (as large as $V_p=16.0 E_r$ -- not shown in the phase diagram), reminiscent of the Arnold tongues found in parametrically driven systems~\cite{McLachlan:1951,Magnus:1966,Molignini:2018}.
In this lobe, the particles undergo density oscillations around their initial pinned configuration, indicating a potential resonance between DDI and quasiperiodic potential 
[see Fig.~\ref{fig:PD-p-3-pos-W}(d) and Fig.~\ref{fig:dens-resonance}(b)-(d)].
While both regions above and below the intermediate phase lobe show strong localization, their behavior is different. 
Above the lobe (high $V_d$, Fig.~\ref{fig:dens-resonance}(e)-(f)), the particles do not move at all from their initial configuration.
Below the lobe (low $V_d$, Fig.~\ref{fig:dens-resonance}(a)), the particle density periodically leaks back and forth from the initial boundaries while retaining a strong localization inside them, indicating a dynamically self-bound state (DSBS).
Such states were recently discovered in bosonic lattice models with attractive DDI~\cite{Barbiero:2015, Li:2020, Li:2021, Li:2021-PRL, Korbmacher:2023}. 
Here, we reveal that DSBS can also exist for fermionic gases with repulsive DDI and away from the lattice limit $V_d \to \infty$.

\begin{figure}[h!]
\centering
\includegraphics[width=\columnwidth]{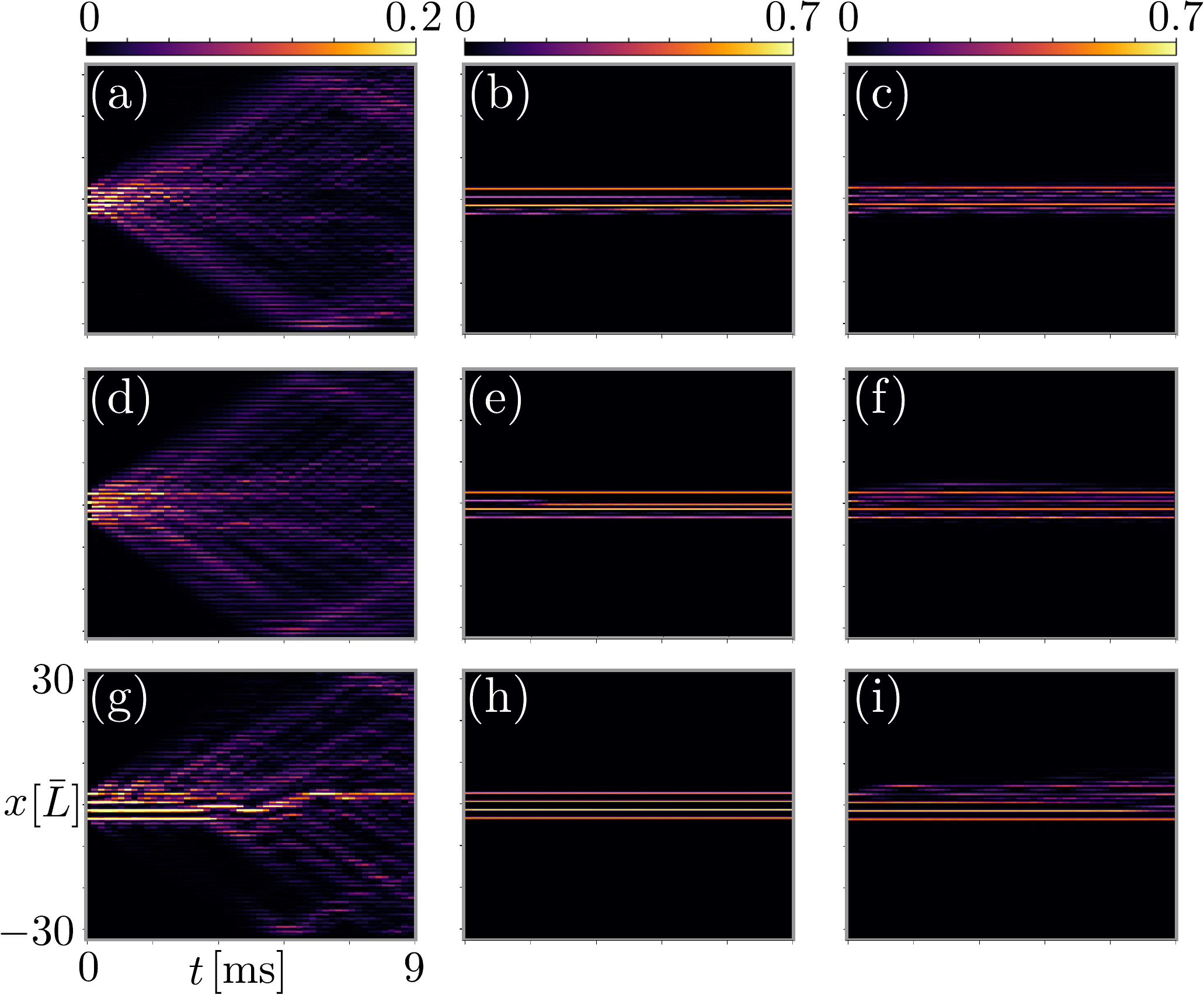}
\caption{
\textbf{Dynamics of dipolar fermions with increasing DDI strength $W$ in a quasiperiodic lattice}.
The interaction strength is (a)-(c) $W=0.02 E_r$, (d)-(f) $W=0.2 E_r$, (g)-(i) $W=2.0 E_r$. 
The potential depths are
(a), (d), (g) $V_p=3.0 E_r$, $V_d=0.0 E_r$ (extended phase)
(b), (e), (h) $V_p=8.0 E_r$, $V_d=1.2 E_r$ (localized phase),
(c) $V_p=7.0 E_r$, $V_d=0.2 E_r$,
(f) $V_p=6.0 E_r$, $V_d=0.4 E_r$,
(i) $V_p=5.0 E_r$, $V_d=0.6 E_r$ (intermediate phase).
The axes in (g) apply to all panels.
}
\label{fig:densities-p-3}
\end{figure}

A more systematic calculation of the density dynamics, shown in Fig.~\ref{fig:densities-p-3}, allows us to better discuss the behavior in each phase.
The dynamics in the extended phase, localized phase, and intermediate phase is pictured respectively in the left, middle, and right columns.
The rows show the dynamics for increasing interaction strength.
In the extended phase [Fig.~\ref{fig:densities-p-3}(a), (d), (g)], the DDI have initially little effect on the density, which maintains its linear spread in time consistent with the Lieb-Robinson bounds~\cite{Tran:2021}.
At strong interactions, however, the long-range repulsive tail in the DDI hinders the spread of correlations~\cite{Bera:2019}, resulting in both a delayed expansion and a more chaotic density spread [Fig.~\ref{fig:densities-p-3}(g)].
In the localized phase [Fig.~\ref{fig:densities-p-3}(b), (e), (h))] very small oscillations can be seen at one edge of the density profile for small and intermediate interaction strengths.
At strong interactions, however, this behavior is suppressed.
This stabilization of localization is also observed in the intermediate phase [Fig.~\ref{fig:densities-p-3}(c), (f), (i)].
At low DDI, the density in the intermediate phase maintains a strong imbalance towards the initial superlattice configuration (due to localized states) while expanding towards outer lattice sites (due to extended states).
However, stronger DDI increase the imbalance while simultaneously reducing the expansion.
This behavior is again consistent with a less effective spread of correlations in dipolar systems due to the interference between the repulsive tail in the interactions.
Furthermore, it is an indication that DDI can be used to control the value of the SPME similarly to what was shown for contact interactions~\cite{Wang:2022}.

\begin{figure}[h!]
\centering
\includegraphics[width=\columnwidth]{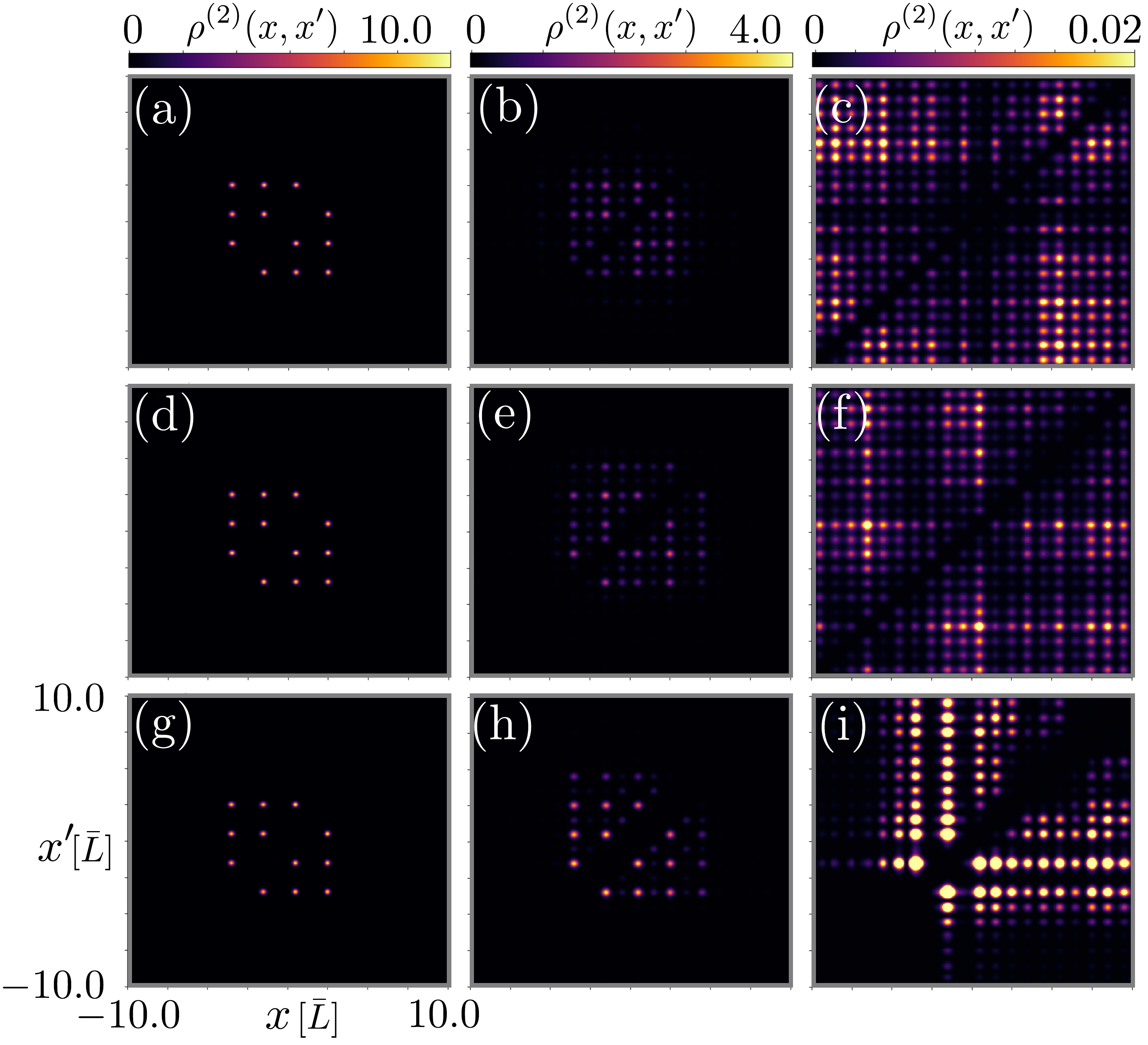}
\caption{
\textbf{2-RDM in the dynamics of fermions with DDI.}
The snapshots are taken at $t=20 \bar{t} \approx 3.6$ ms and for increasing DDI: (a)-(c) $W=0.02 E_r$, (d)-(f) $W=0.2$$E_r$, (g)-(i) $W=2.0 E_r$.
The parameters are 
(a), (d), (g) $V_p=16.0 E_r$, $V_d=2.4 E_r$ (localized phase),
(b), (e), (h) $V_p=5.0 E_r$, $V_d=0.4 E_r$ (intermediate phase),
(c), (f), (i) $V_p=3.0 E_r$, $V_d=0 E_r$ (extended phase).
The axes in (g) apply to all panels.
}
\label{fig:corr-int}
\end{figure}

The increase in localization due to DDI interference is also observed in the 2-RDM.
This is shown in Fig.~\ref{fig:corr-int} for the three different phases and at three increasing values of $W$.
The behavior for weak DDI [$W=0.02 E_r$, Fig.~\ref{fig:corr-int}(a)-(c)] is similar to what already observed in the noninteracting case.
However, when DDI are increased [$W=0.2 E_r$ in Fig.~\ref{fig:corr-int}(d)-(f) and $W=2.0 E_r$ in Fig.~\ref{fig:corr-int}(g)-(i)], the correlation spread becomes progressively stiffer for all phases.
This leads to a complete freeze of correlations in the localized phase [Fig.~\ref{fig:corr-int}(g)], while in the intermediate phase correlations are able to spread only partially~\footnote{We attribute this unidirectional spread to pinning by the given choice of the quasiperiodic potential. Thus, the location of the pinning might change with a different choice of the wave vector ratio and detuning phase.}.
Even for the extended phase, the pinning effect is visible, with correlations spreading more easily towards the right than the left.
This observation agrees with the uneven density spread observed in Fig.~\ref{fig:densities-p-3}(g).

\emph{Discussion} ---
There are several mechanisms at play that explain the overall shape of the phase diagram at strong DDI.
The localized state appearing at large $V_p$ (lattice limit) in the clean limit $V_d \to 0$ is a DSBS stabilized by DDI, 
as evinced by the small density oscillations at the boundaries suggesting dynamical self-confinement [see Fig.~\ref{fig:dens-resonance}(a)].
The $1/r^3$ tails of the DDI cause an interference that locks particles within a certain distance, as demonstrated in similar bosonic systems~\cite{Barbiero:2015, Li:2020, Li:2021, Li:2021-PRL, Korbmacher:2023}, and is likely connected with Hilbert space fragmentation~\cite{Khemani:2020, Herviou:2021, Cheng:2023}.
Away from the clean limit ($V_d \neq 0$, but $V_d \ll V_p$), correlated disorder can be regarded as a perturbation on top of the DSBS.
 In this setting, it is known that the DDI give rise to effective correlated hopping at a distance, i.e. between the edges of the localized clusters~\cite{Levitov:1989, Levitov:1990, Burin:1994,  Burin:2015, Burin:2015-2, Burin:2017, Nosov:2019}, which explains the resonant-like behavior at the edges at their progressive delocalization [Fig.~\ref{fig:dens-resonance}(b)-(d)].
Increasing the disorder further eventually pushes the system again into localization [Fig.~\ref{fig:dens-resonance}(f)].
In this limit, the system approaches an Aubry-Andr\'{e} model, whose localized phase is known to be stable against DDI~\cite{Gopalakrishnan:2017, Nag:2019, Deng:2019}.

On the other hand, as $V_p$ is decreased, the lattice becomes shallower, increasing the effective particle hopping.
In the clean limit ($V_d \to 0$), this progressively reduces the critical distance under which particles can be confined by the DDI~\cite{Barbiero:2015} and full localization is lost.
With correlated disorder ($V_d \gg 0$), the loss of localization is analogous to what happens in the noninteracting case [cf. Fig.~\ref{fig:pd}(a)].
However, the repulsive nature of the DDI -- unhindered by the DSBS resonances in shallow lattices -- provides more mobility, resulting in an enlarged intermediate phase.

Repulsive composite objects have been shown to be stable with respect to on-site interactions in a structured environment such as a simple optical lattice~\cite{Winkler:2006,Strohmaier:2010,Schneider:2012,Ronzheimer:2013,Zaletel:2015}.
Our study propels these notions to a full many-body horizon by combining quasiperiodicity and dipolar interactions in a realistic quantum simulator setting~\cite{Hughes:2023,Bilinskaya:2024}.
Our scheme, inspired by experiments in 1D geometries, could be readily extended to higher-dimensional setups where a richer landscape of quasicrystalline phases with different rotational symmetries can be engineered~\cite{Viebahn:2019, Sbroscia:2020, Gottlob:2023}.
Moreover, clean system localization can appear when other mechanisms induce effective disorder, such as an external linear field~\cite{Schulz:2019, Yao:2020, Taylor:2020, Doggen:2021, Guo:2021, Yao:2021}, slow interaction dynamics~\cite{Yao:2016}, or both~\cite{vanNieuwenburg:2019}.
Given that the system presented in our work interpolates between the clean and the quasiperiodic case, it should help shed light on commonalities across a multitude of exotic localization phenomena that combine long-range interactions and correlated disorder.


\emph{Acknowledgments} --- We acknowledge computation time on the ETH Zurich Euler cluster, at the High-Performance Computing Center Stuttgart (HLRS), and on the SUNRISE cluster of Stockholm University.
We thank O. Alon, E. Bergholtz, R. Chitra, E. Fasshauer, I. Khaymovich, and R. Lin for useful discussions.
This work is supported by the Swedish Research Council (grants 2018-00313 and 2024-05213) and Knut and Alice Wallenberg Foundation (KAW) via the project Dynamic Quantum Matter (2019.0068).

\bibliography{ultracold}

\clearpage
\pagebreak
\widetext
\begin{center}
\textbf{\large Supplemental Material for \\ Stability of quasicrystalline ultracold fermions to dipolar interactions}
\end{center}
\setcounter{equation}{0}
\setcounter{figure}{0}
\setcounter{table}{0}
\setcounter{page}{1}
\makeatletter
\renewcommand{\theequation}{S\arabic{equation}}
\renewcommand{\thefigure}{S\arabic{figure}}

\section{System parameters}
\label{app:parameters}
Throughout the main text, unless otherwise stated, we perform simulations with $N=4$ fermions and $M=9$ orbitals.
The quasiperiodic optical lattice consists of a superposition of two potentials: a primary lattice of depth $V_p$ and wavelength $\lambda_p$, and a detuning lattice of depth $V_d$, wavelength $\lambda_d$, and phase $\phi$:
\begin{equation}
V(x) = \frac{V_p}{2} \cos (2 k_p x) + \frac{V_d}{2} \cos(2 k_d x + \phi).
\end{equation}
where $k_i = \frac{2\pi}{\lambda_i}$.
We choose the wavelengths to be compatible with real experimental realizations in ultracold atomic labs, i.e. $\lambda_p \approx 532.2$ nm and $\lambda_d \approx 738.2$ nm.
This gives wave vectors $k_p \approx 1.1806 \times 10^7$ m$^{-1}$, $k_d \approx 8.5115 \times 10^6$ m$^{-1}$. 
For the figures in the main text, we chose $\phi = 4.0$.
However, we performed calculations with various random value of $\phi$ and did not notice any qualitative difference in the dynamics.

\subsection{Lengths}
In MCTDH-X simulations, we choose to set the unit of length $\bar{L} \equiv \frac{\lambda_p}{2} = 266.1$ nm, which makes the maxima of the primary lattice appear at integer values in dimensionless units.
We run simulations with 2'048 gridpoints in an interval $x \in [-32 \bar{L}, 32 \bar{L}] \approx [-8.5152 \mu\mathrm{m}, 8.5152 \mu\mathrm{m}]$, giving a resolution of around 8.3156 nm.
We employ hard-wall boundary conditions at the end of the probe spatial interval.

\subsection{Energies}
The unit of energy is defined in terms of the recoil energy of the primary lattice, i.e. $E_r \equiv \frac{\hbar^2 k_p^2}{2m} \approx 2.89 \times 10^{-30}$ J with $m \approx 2.673$ 10$^{-25}$ kg the mass of $^{161}$Dy atoms.
The corresponding recoil frequency is $\nu_r = E_r/h \approx 4.36 kHz$.
More specifically, we define the unit of energy as 
$\bar{E} \equiv \frac{\hbar^2}{m} \frac{1}{\bar{L}^2} = \frac{2E_r}{\pi^2} \approx 5.86 \times 10^{-31}$ J, or in frequency, $\nu_{\bar{E}}=884$ Hz.

In typical experiments with quasiperiodic optical lattices, the depths are varied in regimes of up to around 8 recoil energies for the primary lattice, and around 1 recoil energy for the detuning lattice.
In our simulations, we probe similar regimes:
$V_p \in [15 \bar{E}, 60 \bar{E}] \approx [3.0 E_r,  12.2 E_r]$, $V_d \in [0 \bar{E}, 6 \bar{E}] = [0 E_r,  1.2 E_r]$.
In terms of frequencies, 
$V_p \in [13.1, 53.2]$ kHz, 
$V_d \in [0, 5.2]$ kHz.

When we turn on dipolar interactions, we probe the regimes with strength $W \in [0 \bar{E}, 10.0 \bar{E}] \approx [0 E_r, 2 E_r]$ (in terms of frequencies: $W \in [0, 8.7] kHz$), which is in accordance to the values that can be achieved in near-term ultracold dipolar quantum simulators.

\subsection{Time}
The unit of time is also defined from the unit of length as 
$\bar{t} \equiv \frac{m \hat{L}^2}{\hbar} = \frac{m \lambda_p^2}{4 \hbar} = 179.4$ $\mu$s,
or in frequency terms
$\nu_{\bar{t}} = \frac{1}{\bar{t}} = 5.57$ kHz.
In our simulations, we run time evolutions up until around $t \approx 250 \bar{t} \approx 44.85$ ms.
Note that this is much shorter compared to experimental runs (in the order of a couple of seconds), but it allows us to probe all the essential features of the system because its dynamics is quite fast.
In fact, the core features of the phase diagram can be faithfully extracted already at $t = 50 \bar{t} \approx 8.97$ ms.
Furthermore, since are dealing with a smaller system size compared to the experiments, we need to extract expansion measurements at a much shorter time scale, before the particles hit the boundaries of the system.

\subsection{Simulation time}
The number of configurations in the MCTDH ansatz can be taken as a proxy of the complexity of MCTDH-X calculations and there are $\left( \begin{array}{c} M \\ N  \end{array} \right) \approx \frac{M^N}{N!}$ configurations for $N$ particles in $M$ orbitals. 
As a result, calculations with larger $N$ and $M$ will be exponentially slower.
In table \ref{table:times}, we report the average final times (in ms) used in the simulations reported in this work for different values of $N$ and $M$.
The simulations ran for 5 days on the ETH Euler supercomputing cluster (mainly on AMD EPYC processors with average 2.4 GHz nominal speed).

\begin{table}[h!]
\centering
\begin{tabular}{|| c | c | c ||}
\hline \hline
N & M & average time [ms] \\ \hline \hline
4 & 6 & 130.8 \\ \hline
4 & 8 & 52.6 \\ \hline
6 & 8 & 55.5 \\ \hline
6 & 10 & 27.8 \\
\hline \hline
\end{tabular}
\caption{Average final simulation times for the dynamics of $N$ fermions and $M$ orbitals in a quasiperiodic optical lattice obtained on 2.4 GHz processors for runs of 5 days.}
\label{table:times}
\end{table}

\newpage
\section{MCTDH-X theory}
The MCTDH-X software numerically solves the many-body Schr\"{o}dinger equation for a given many-body Hamiltonian describing $N$ interacting, indistinguishable bosons or fermions subject to a one-body potential.
It is based on the MCTDH ansatz for the wavefunction, i.e. a time-dependent superposition of time-dependent many-body basis functions:
\begin{eqnarray}
\vert \Psi(t) \rangle &=& \sum_{\vec{n}} C_{\vec{n}}(t) \vert \vec{n}; t \rangle; \;\; \vec{n}=\left(n_1,...,n_M\right)^T;\nonumber \\
\vert \vec{n}; t \rangle &=& \mathcal{N}  \prod_{i=1}^M \left[ \hat{b}_i^\dagger(t) \right]^{n_i} \vert \text{vac} \rangle; \qquad \phi_j(\mathbf{x};t)=\langle \mathbf{x} \vert \hat{b}_j (t) \vert 0 \rangle.  
\label{eq:ansatz}
\end{eqnarray}
Here, the $C_{\vec{n}}(t)$ are referred to as coefficients, the $\vert \vec{n}; t \rangle$ as configurations, and the normalization factor is  $\mathcal{N}=\frac{1}{\sqrt{\prod_{i=1}^{M} n_i!}}$ for bosons and $\mathcal{N}=1$ for fermions. 
Each configuration is constructed from $M$ orthonormal time-dependent single-particle functions, or \emph{orbitals}, $\lbrace \phi_k(\mathbf{x},t); k=1,...,M \rbrace$, and is chosen to respect the underlying particle statistics (fully symmetric for bosons and fully anti-symmetric for fermions).
For fermions, $M>N$ orbitals are required due to the Pauli principle.
By minimizing the action obtained from the many-body Hamiltonian rewritten with the MCTDH ansatz, we obtain a set of equations of motion for the parameters in Eq.~\eqref{eq:ansatz}.
These are a set of coupled first-order differential equations for time-dependent coefficients $C_{\vec{n}}(t)$ and non-linear integro-differential equations for the orbitals  $\phi_j(\mathbf{x};t)$.
The MCTDH-X software can integrate these equations of motion either in imaginary time (for time-independent Hamiltonians) to obtain many-body ground-state properties, or in real time to perform full-time propagation.

\newpage
\section{Orbital convergence}
The MCTDH-X ansatz becomes numerically exact in the limit $M \to \infty$. 
However, very often a finite number of orbitals is enough to accurately describe the ground-state properties and the short-to-medium dynamics.
The population of each orbital is called \emph{occupation} $\rho_i$, and is defined as the eigenvalues of the reduced one-body density matrix,
\begin{equation}
\rho^{(1)}(x, x') = \frac{1}{N} \left< \Psi \right| \hat{\Psi}^{\dagger}(x') \hat{\Psi}(x) \left| \Psi \right> = \sum_{i} \rho_i \left( \phi_i^{\mathrm{(NO)}}(x') \right)^* \phi_i^{\mathrm{(NO)}}(x),
\end{equation}
where the eigenvectors $\phi_i^{\mathrm{(NO)}}(x)$ are termed natural orbitals.
The occupations are ranked from most occupied to least occupied, and are normalized such that they sum up to one.
A measure of the exactness of the MCTDH-X ansatz is given by \emph{orbital convergence}: an MCTDH-X calculation with $M$ orbitals can be declared converged in orbitals number within a given threshold $C$, if the population of an additional orbital stays within that threshold, i.e. $\rho_{M+1} < t$.
Typical threshold values are $C \gg \frac{1}{M}$, e.g. for $M=9$, $t=0.01$.

\begin{figure}[h!]
\centering
\includegraphics[width=0.7\columnwidth]{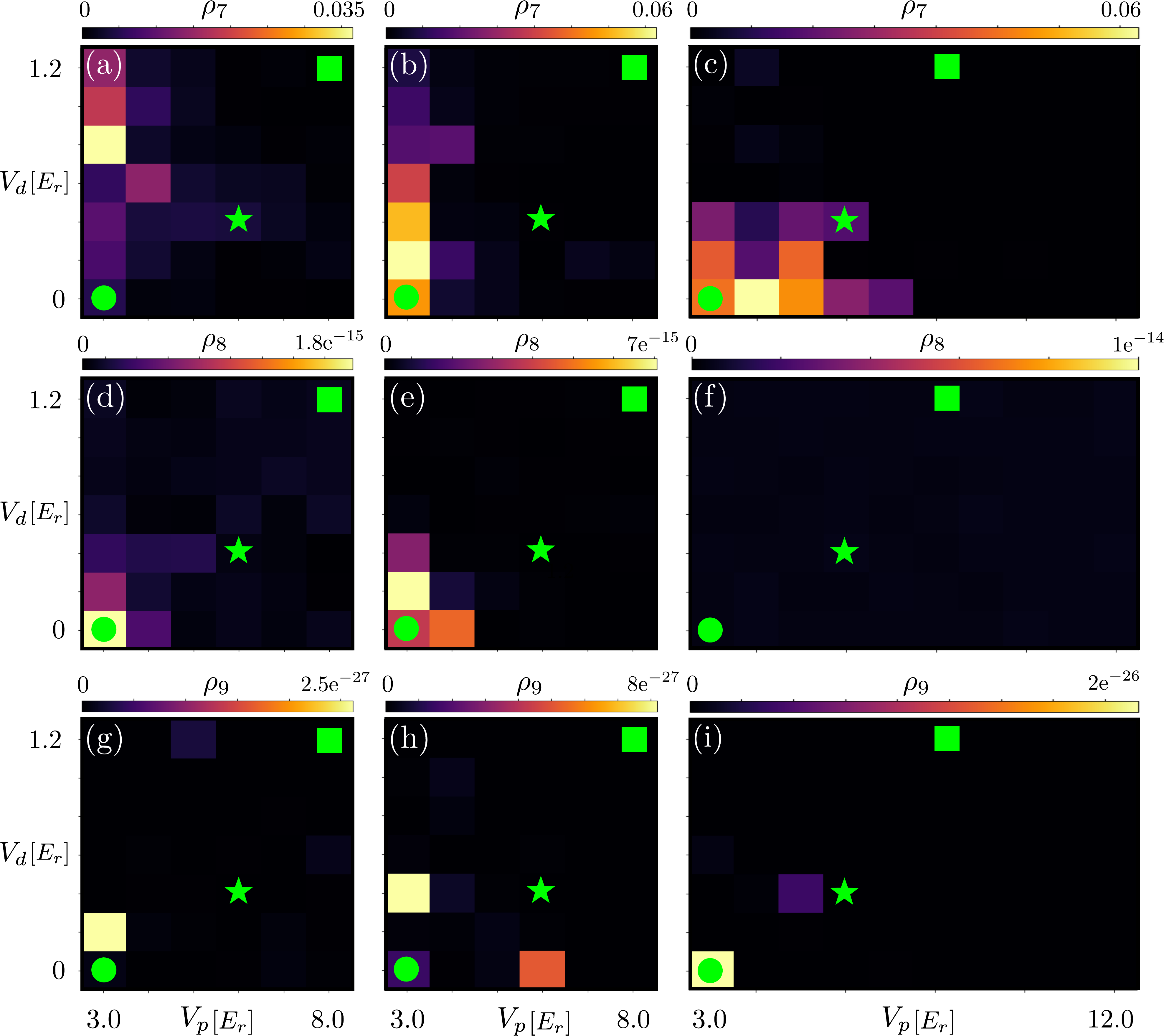}
\caption{
Maximal occupation of the three least occupied orbitals ((a)-(c): third-least occupied, (d)-(f) second-least occupied, (g)-(i): least occupied), during the time evolution, as a function of $V_p$, $V_d$, and $W$.
The green symbols indicate the points in the three different phases shown in Fig.~\ref{fig:occ-time}: extended phase (circle), intermediate phase (star), and localized phase (square).
The different columns correspond to different values of interaction strength $W$: 
(a),(d),(g): $W=0.02 E_r$,
(b),(e),(h): $W=0.2 E_r$,
(c),(f),(i): $W=2.0 E_r$.
}
\label{fig:occ-pd}
\end{figure}

In the interacting calculations reported in the main text, we have used $N=4$ and $M=9$.
For the noninteracting case, using $M=8$ was sufficient because the population of the last \emph{four} orbitals is negligible for the entire parameter space.
Using $M=9$ in the interacting case gives us the best compromise between increasing computational complexity (which scales exponentially with the number of orbitals) and achieving orbital convergence.
In Fig.~\ref{fig:occ-pd}, we report the maximal occupation value of the three least populated orbitals (i.e. the seventh, eighth, and ninth orbital) for each time evolution in parameter space.
As we can see, the occupations of the eighth and ninth orbitals are insignificant through the entire parameter space. 

\begin{figure}[h!]
\centering
\includegraphics[width=0.6\columnwidth]{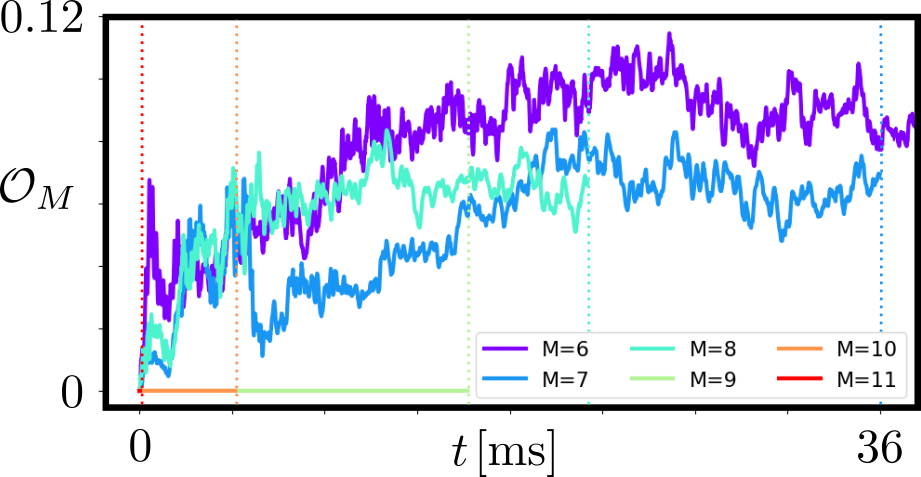}
\caption{
Maximal occupation of the least occupied orbital during the time evolution for simulations with increasing number of orbitals $M$.
The point at $V_p=3.0 E_r$, $V_d=0 E_r$ and interaction strength $W=2.0 E_r$ was chosen as representative.
The calculations were let run for 5 days on AMD EPYC processors with average 2.4 GHz nominal speed.
The vertical dotted lines indicate the maximal time step reached by each computation with different $M$.
}
\label{fig:comparison-occ-M}
\end{figure}

In Fig.~\ref{fig:comparison-occ-M}, we compare the occupation of the least occupied orbitals during the time evolution with different number of orbitals.
The point at $V_p=3.0 E_r$, $V_d=0 E_r$ (belonging to the extended phase) and interaction strength $W=2.0 E_r$ was chosen as representative since, empirically, we find that the extended phase is the region in parameter space that requires the highest number of orbitals for a converged description of the correct dynamics.
We can see that up until $M=8$, the least occupied orbitals still retains a macroscopic population.
However, from $M=9$ onwards, the least occupied orbital acquires a negligible occupation (and actually -- as seen from Fig.~\ref{fig:occ-pd} -- this also applies to the second least occupied orbital).
The calculations with $M=10$ and higher are however much slower and can only reach a fraction of the final time reached with $M=9$.
This justifies our selection of $M=9$ for the main text calculations.

\begin{figure}[h!]
\centering
\includegraphics[width=0.8\columnwidth]{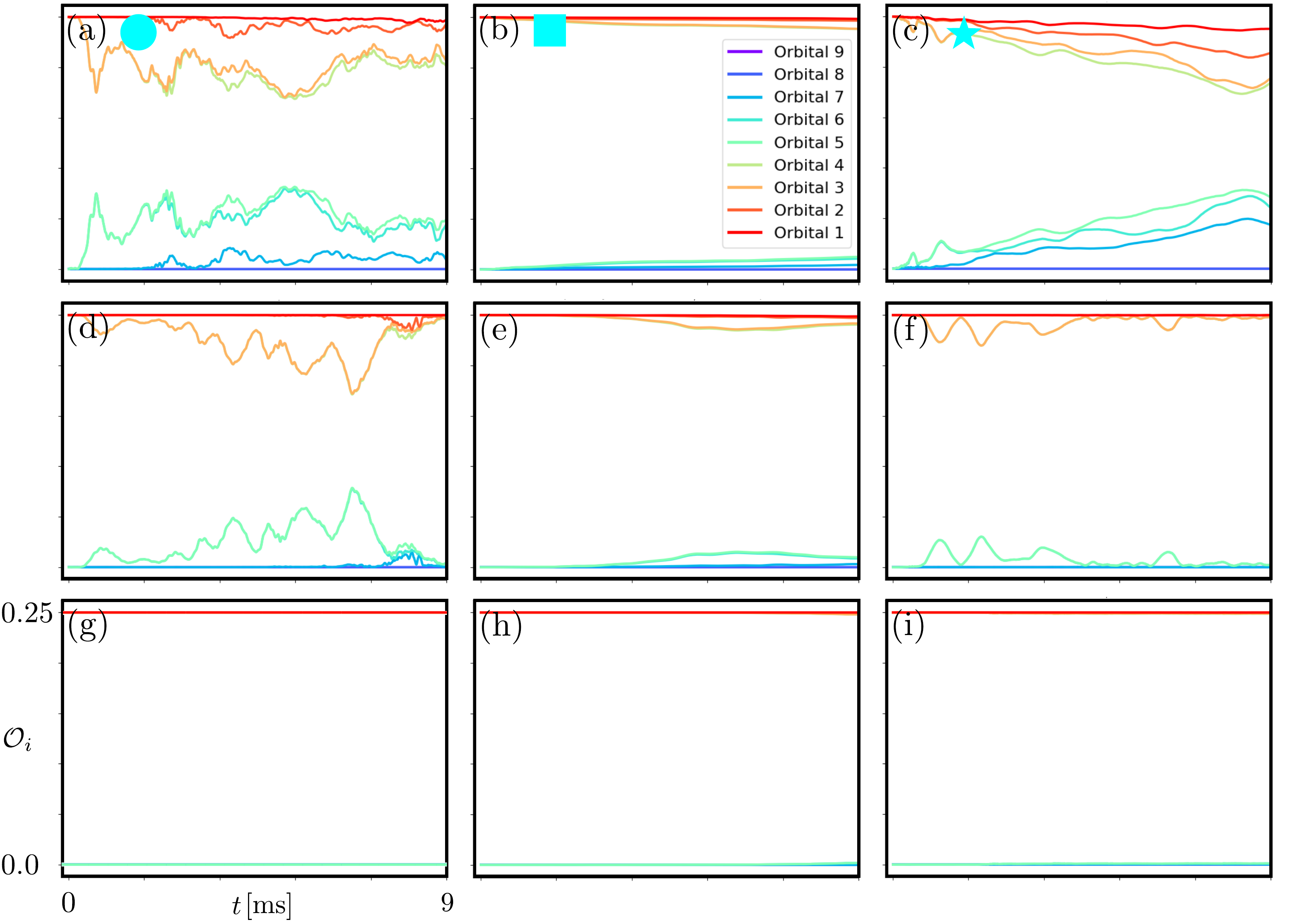}
\caption{
Time evolution of orbital occupations in each phase for increasing interaction strength $W$: extended phase (left column), localized phase (middle column), and intermediate phase (right column).
The interaction strength is (a)-(c) $W=0.02$$E_r$, (d)-(f) $W=0.2$$E_r$, (g)-(i) $W=2.0$$E_r$. 
The potential depths are
(a), (d), (g): $V_p=3.0$$E_r$, $V_d=0.0$$E_r$,
(b), (e), (h): $V_p=8.1$$E_r$, $V_d=1.2$$E_r$,
(c), (f), (i): $V_p=6.1$$E_r$, $V_d=0.4$$E_r$.
The symbols indicate the phase of each column and where in parameter space they are taken from (cf. Fig.~\ref{fig:occ-pd}): extended phase (circle), intermediate phase (star), and localized phase (square).
}
\label{fig:occ-time}
\end{figure}

Finally, in Fig.~\ref{fig:occ-time}, we show the full time evolution of the orbital occupation at a representative point in each phase (extended, intermediate, and localized) for the same three values of the interaction strength $W$ shown in Fig.~\ref{fig:occ-pd}.
From the dynamics we can once again see that the extended phase generally requires a higher population in more orbitals, whereas in the intermediate and localized phases there is a clear dominance of the first four orbitals.
These results indicate that our simulations are fully converged in the number of orbitals for each point in parameter space and should be accurate enough to guarantee a correct representation of the dynamics in the probed time regimes.

\newpage
{\color{white}.}
\newpage
{\color{white}.}
\newpage
\section{Orbital profiles}
In this section, we visualize the profiles of the $M=9$ orbitals used to calculate the dynamics of $N=4$ particles presented in the main text.
This visualization should help the reader understand that the orbitals are truly optimized both in space and in time, and might end up acquiring a very nonlocal profile when the optical lattices are shallow and the lattice description is a very coarse approximation.

\begin{figure}[h!]
\centering
\includegraphics[width=0.8\columnwidth]{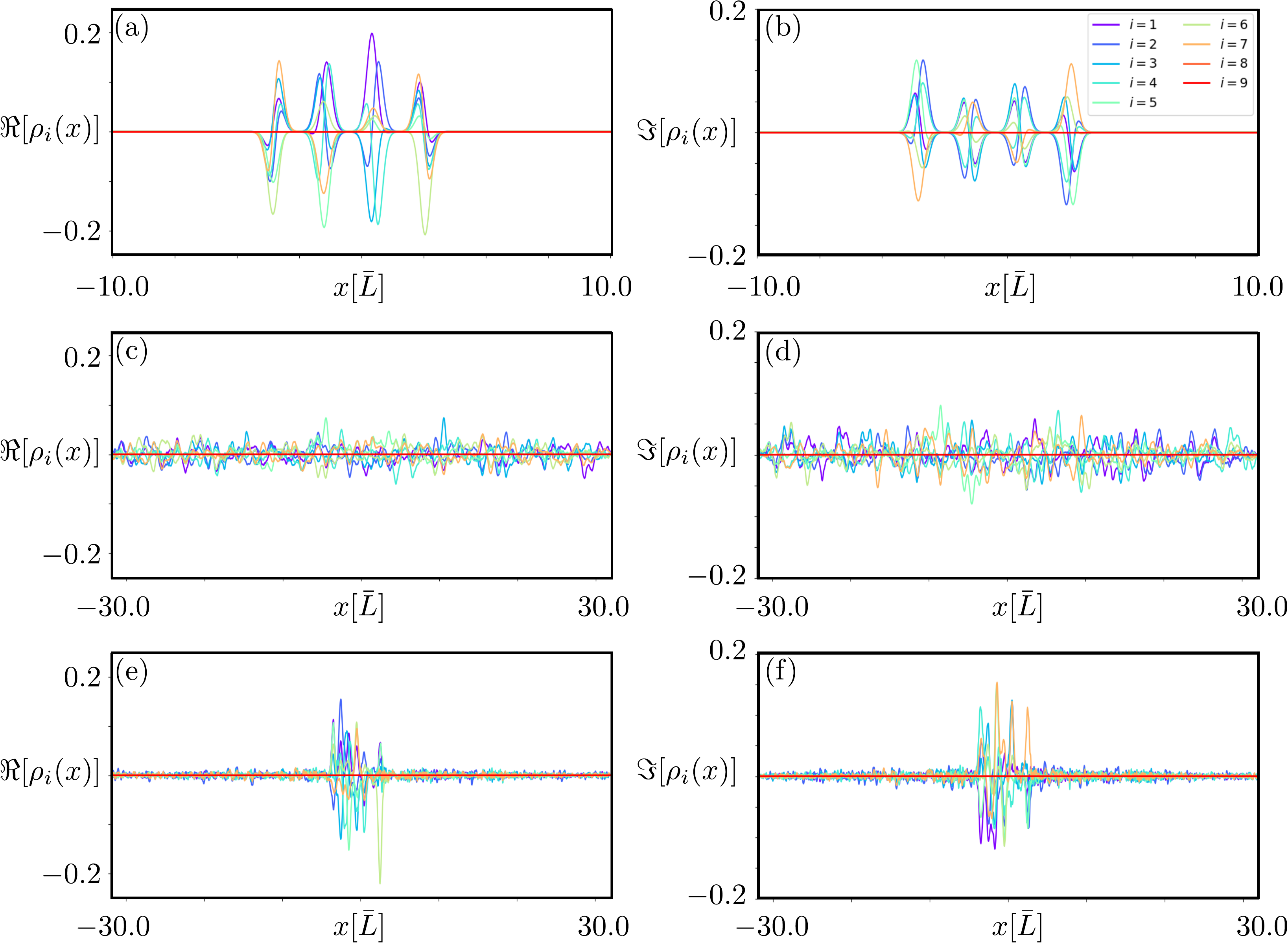}
\caption{
Examples of real (left columns) and imaginary (right columns) amplitudes for the MCTDH orbitals used in the many-body expansion. 
The index $i$ labels the orbitals from largest to smallest contribution.
(a)-(b) Orbital profiles for the initial state at $t=0$.
(c)-(d) Orbital profiles at time $t=50 \bar{t}$ for the system in the extended phase with $V_p = 3.0 E_r$, $V_d=0.4 E_r$, $W=0.02 E_r$.
(e)-(f) Orbital profiles at time $t=50 \bar{t}$ for the system in the localized phase with $V_p = 8.0 E_r$, $V_d=1.2 E_r$, $W=0.02 E_r$.
}
\label{fig:orbitals}
\end{figure}

Exemplary orbitals are depicted in Fig.~\ref{fig:orbitals}. 
Panels (a)-(b) visualize the orbitals at time $t=0.0$ (their profile is identical for all phases since they all start from the same initial state in the time evolution) -- obtained from imaginary time evolution into the ground state corresponding to four sites in the center of the optical lattice.
Already here we can appreciate that the orbitals are delocalized over many sites. 
This freedom of delocalization is what gives MCTDH orbital a high degree of expressivity.
Panels (c)-(d) show the orbitals at time $t=50 \bar{t}$ for a data point in the extended phase, while panel (e)-(f) shows the orbitals at time $t=50 \bar{t}$ for a data point in the localized phase (note the different scale in the $x$-axis).
As we can see from the plots, in the extended phase the orbitals become highly nonlocal during time evolution, to follow the density expansion.
In the localized phase, instead, the orbitals retain most of their weight in the center of the system.

\newpage
\section{Results for different particle and orbital number}
\begin{figure}[h!]
\centering
\includegraphics[width=0.5\columnwidth]{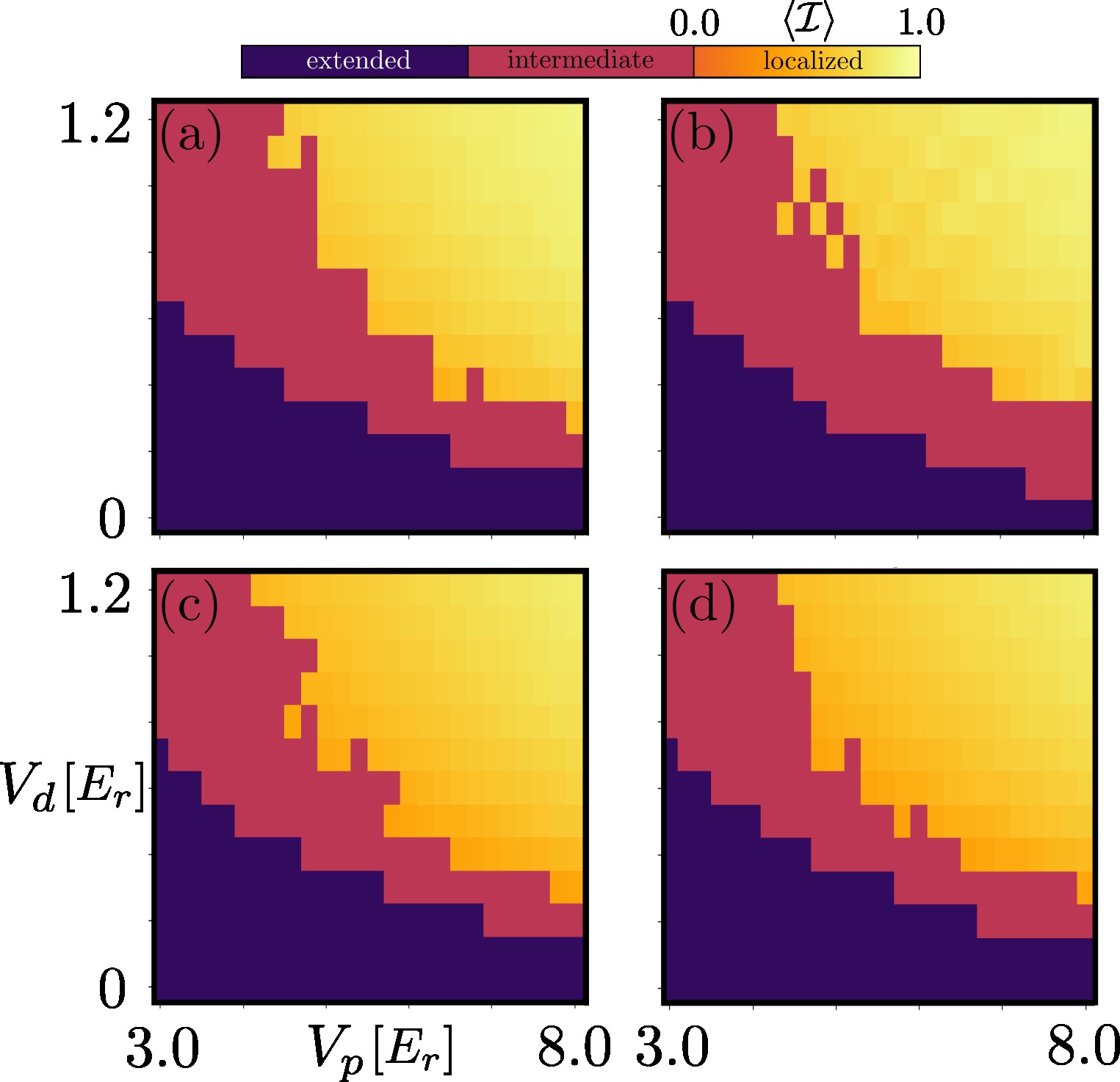}
\caption{
Phase diagrams for $N$ ultracold fermions in a 1D quasiperiodic potential.
(a) $N=4$ particles, $M=6$ orbitals. 
(b) $N=4$ particles, $M=8$ orbitals. 
(c) $N=6$ particles, $M=8$ orbitals. 
(d) $N=6$ particles, $M=10$ orbitals. 
For the localized phase, the time-averaged imbalance parameter $\left< \mathcal{I} \right>$ is further plotted as a continuous color scheme.
}
\label{fig:pd-N}
\end{figure}

In this section, we show that the noninteracting phase diagram does not change qualitatively when changing the  numbers of particles ($N=4,6$) and orbitals ($M=6,8$).
Our results are shown in Fig.~\ref{fig:pd-N}, where the phase diagrams are constructed by calculating the imbalance and expansion observables as the depth of the primary and detuning lattices are varied, precisely like in the main text figures.
The parameter ranges are $V_p \in [3.0 E_r, 8.0 E_r]$ and $V_d \in [0, 1.2 E_r]$ as in the main text.
All the other computational parameters (imbalance target time, thresholds for imbalance and edge density etc.) are kept the same as in the main text.

In all cases, we can clearly distinguish the three expected phases.
In particular, increasing the number of orbitals at fix particle number does not dramatically alter the phase boundaries and illustrates that the physical phenomenology can be well described with only $M=N+2$ orbitals.
Changing the number of particles also has negligible qualitative effects and in particular the overall shape of the intermediate phase is preserved.
We do however observe a very slight thinning of this phase for larger particle numbers, which is compatible with experimental observations as experiments are typically carried out with a much larger number of particles.

\newpage
\section{Effective nearest-neighbor hopping}

In this section, we estimate the nearest-neighbor hopping amplitude \( J_{\mathrm{NN}} \) in the lattice regime we simulate using a semiclassical approximation based on localized Wannier functions. 
Although our simulations are performed in the continuum and the primary optical lattice is rather shallow (typically \( V_p \in [3 E_r, 8 E_r] \)), it is still instructive to compare with the tight-binding regime for orientation.

In deep lattice limits, the lowest-band Wannier functions can be approximated by Gaussians, and the hopping between adjacent sites can be estimated analytically. 
We follow the standard approximation (see e.g. Morsch and Oberthaler, Rev. Mod. Phys. \textbf{78}, 179 (2006)):
\begin{equation}
J_{\mathrm{NN}} \approx \frac{4}{\sqrt{\pi}} E_r \left( \frac{V_p}{E_r} \right)^{3/4} \exp\left( -2 \sqrt{\frac{V_p}{E_r}} \right).
\label{eq:JNN}
\end{equation}
Here, \( V_p \) is the depth of the primary optical lattice, and \( E_r \) is the recoil energy defined in the sections above.

To illustrate the scaling, let us consider a few representative lattice depths:
\begin{itemize}
\item For \( V_p = 3 E_r \): \\
\( J_{\mathrm{NN}} \approx 0.142 E_r \Rightarrow J_{\mathrm{NN}} / h \approx 590\,\mathrm{Hz} \), \quad \( \hbar / J_{\mathrm{NN}} \approx 1.69\,\mathrm{ms} \)
\item For \( V_p = 6 E_r \): \\
\( J_{\mathrm{NN}} \approx 0.0645 E_r \Rightarrow J_{\mathrm{NN}} / h \approx 268\,\mathrm{Hz} \), \quad \( \hbar / J_{\mathrm{NN}} \approx 3.73\,\mathrm{ms} \)
\item For \( V_p = 9 E_r \): \\
\( J_{\mathrm{NN}} \approx 0.030 E_r \Rightarrow J_{\mathrm{NN}} / h \approx 125\,\mathrm{Hz} \), \quad \( \hbar / J_{\mathrm{NN}} \approx 8.00\,\mathrm{ms} \)
\end{itemize}
These results show that the hopping times remain much smaller than the maximum time scale that we probe in our numerics (up to 45 ms).

Furthermore, we compare these hopping energies with the strength of dipolar interactions used in our simulations. 
For example, in the strongly interacting case \( V_d = 2 E_r \), we find:
\[
\frac{V_d}{J_{\mathrm{NN}}} \approx \frac{2.0}{0.0645} \approx 31.
\]
Thus, the dipolar interaction energy scale dominates over the kinetic energy scale set by \( J_{\mathrm{NN}} \), confirming that we operate deep in the strongly interacting regime. 
This supports our conclusion that the behavior observed in our simulations -- especially in the localized regime -- cannot be captured perturbatively from a weakly interacting tight-binding perspective and continuum formulations like MCTDH-X become necessary.

\newpage
\section{Dynamics of observables}
In this section, we visualize the full time evolution of the three main observables mentioned in the main text: the imbalance $\mathcal{I}$, the edge density $\mathcal{D}$, and the expansion $\mathcal{E}$.

\begin{figure}[h!]
\centering
\includegraphics[width=0.7\columnwidth]{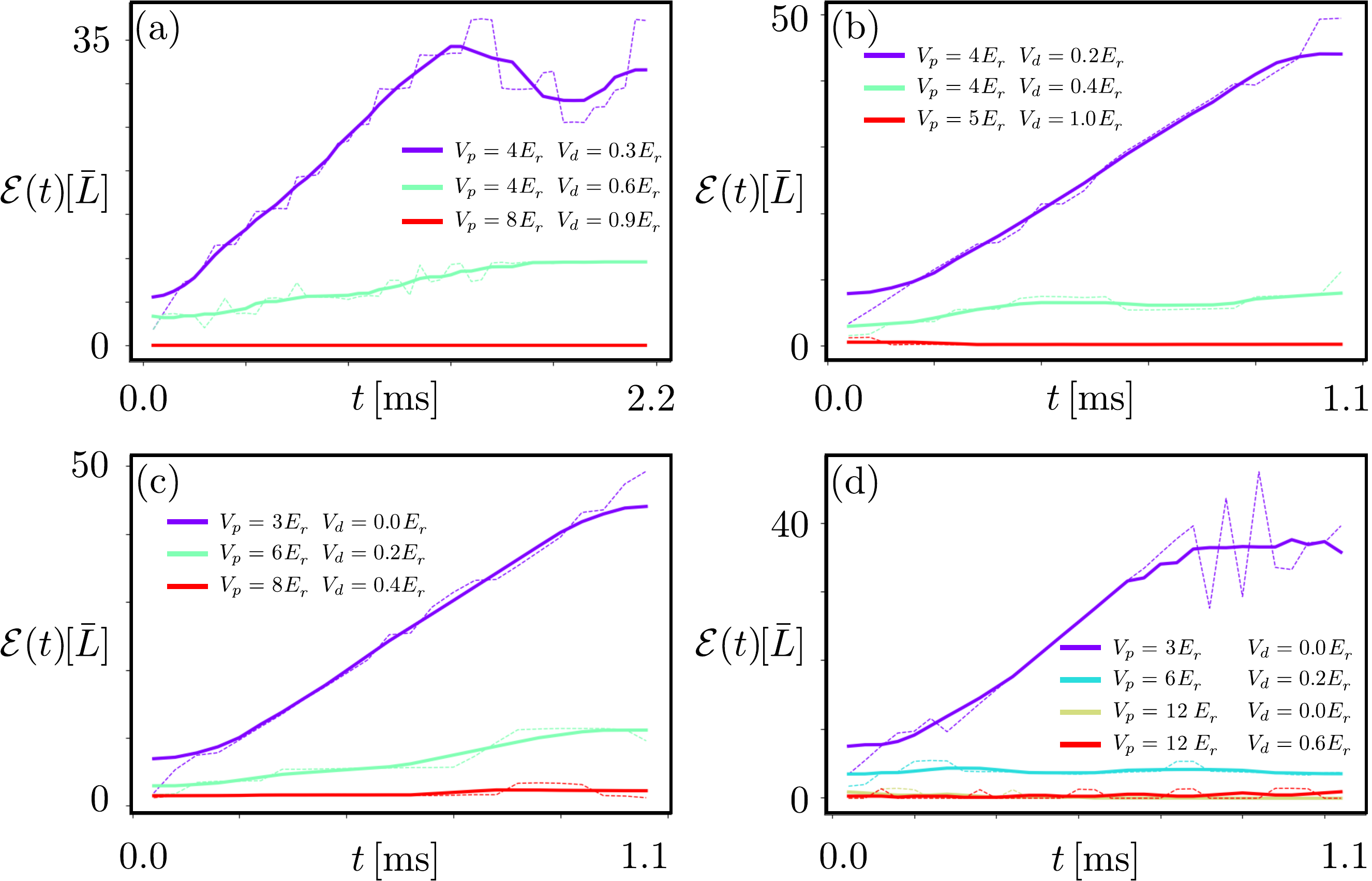}
\caption{
Density expansion dynamics for three different parameter values corresponding to the extended phase (purple curves), the intermediate phase (cyan/blue curves) and the localized phases (olive curve: DSBS state, red curve: standard localized phase).
The different panels depict the dynamics for different values of dipolar interactions: (a) $W=0.0 E_r$, (b) $W=0.02 E_r$, (c) $W=0.2 E_r$, (d) $W=2.0 E_r$.
The dashed lines are the raw data, while the thick solid lines are running averages.
}
\label{fig:expansion}
\end{figure}

Each panel of Fig.~\ref{fig:expansion} shows expansion dynamics for three or four representative parameter choices in the extended, intermediate, and localized phase, respectively, and for a different value of dipolar interaction strength: (a) $W=0.0 E_r$, (b) $W=0.02 E_r$, (c) $W=0.2 E_r$, (d) $W=2.0 E_r$.
To plot all curves in the same panels, and since the expansion is quite rapid in the extended phase, we only show times up to $t=2.2 ms$.
We can clearly distinguish the three qualitatively different behaviors captured by the three phases presented in the main text.
In the extended phase, the density rapidly expands to reach the system boundaries.
In the localized phase, no expansion is observed (we have verified that the expansion remains negligible for longer times).
In the intermediate phase, expansion occurs but it is much slower than in the extended phase, and does not reach the system boundaries for the time scales probed by our numerics.

\begin{figure}[h!]
\centering
\includegraphics[width=0.7\columnwidth]{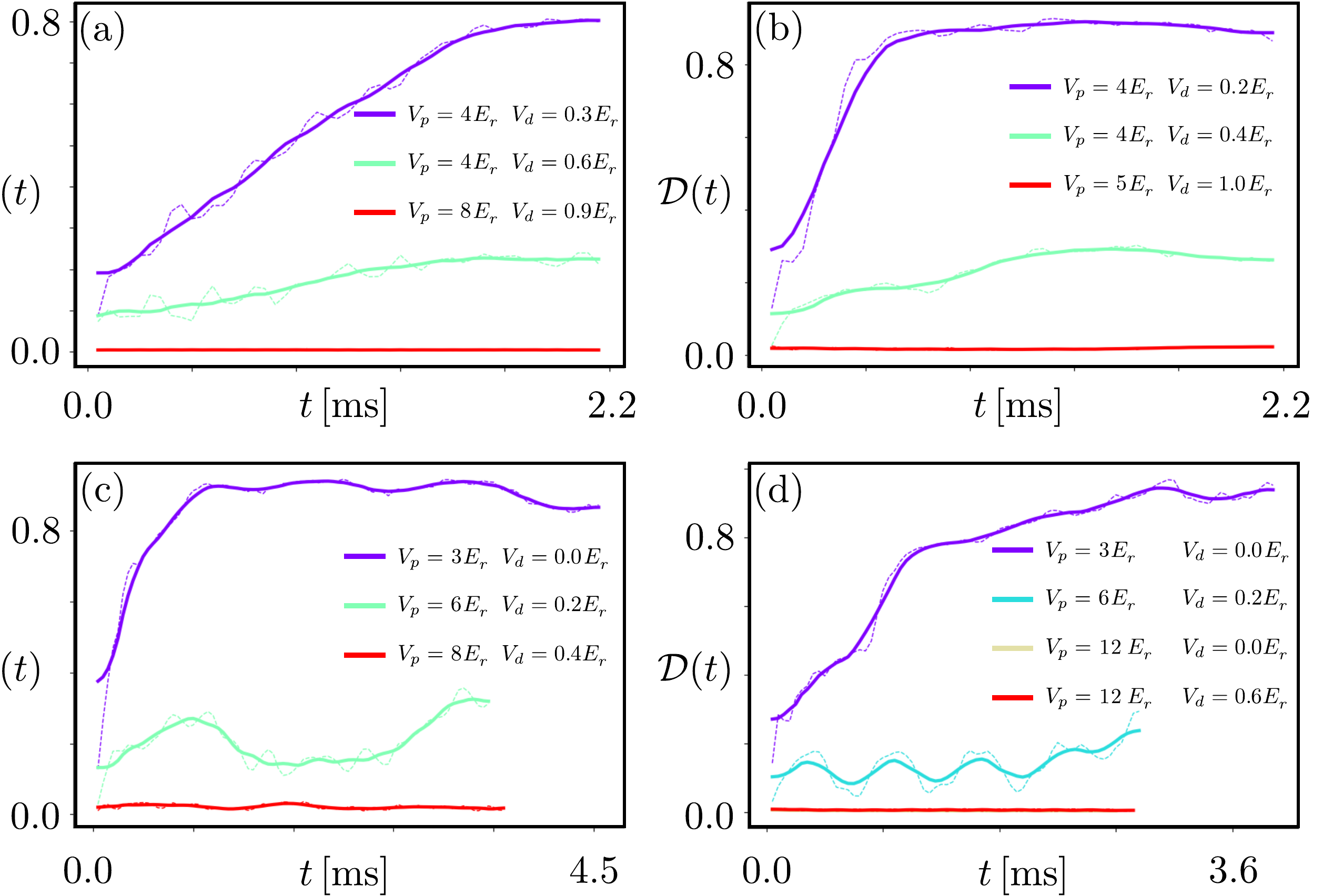}
\caption{Edge density dynamics for three different parameter values corresponding to the extended phase (purple curves), the intermediate phase (cyan/blue curves) and the localized phases (olive curve: DSBS state, red curve: standard localized phase).
The different panels depict the dynamics for different values of dipolar interactions: (a) $W=0.0 E_r$, (b) $W=0.02 E_r$, (c) $W=0.2 E_r$, (d) $W=2.0 E_r$.
The dashed lines are the raw data, while the thick solid lines are running averages.
}
\label{fig:edge-density}
\end{figure}

Fig.~\ref{fig:edge-density} depicts the behavior of the edge density, another expansion proxy which provides a numerically more stable quantity to track the amount of expansion (and allows us to plot different phases on the same scale for longer times).
Each panel shows three or four representative parameter choices in the extended, intermediate, and localized phase, respectively, and for a different value of dipolar interaction strength: (a) $W=0.0 E_r$, (b) $W=0.02 E_r$, (c) $W=0.2 E_r$, (d) $W=2.0 E_r$.
The behavior of the three different phases is similar to that of the raw expansion.
iI the extended phase, the edge density increases fast to a saturation value.
In the localized phase, the edge density stays essentially at zero through time evolution.
In the intermediate phase, the edge density tends to increase, but a slower rate than in the extended phase.
Note also how stronger interactions tend to inhibit the growth rate of the edge density, consistent with the localizing role of strong dipolar repulsions.

\begin{figure}[h!]
\centering
\includegraphics[width=0.7\columnwidth]{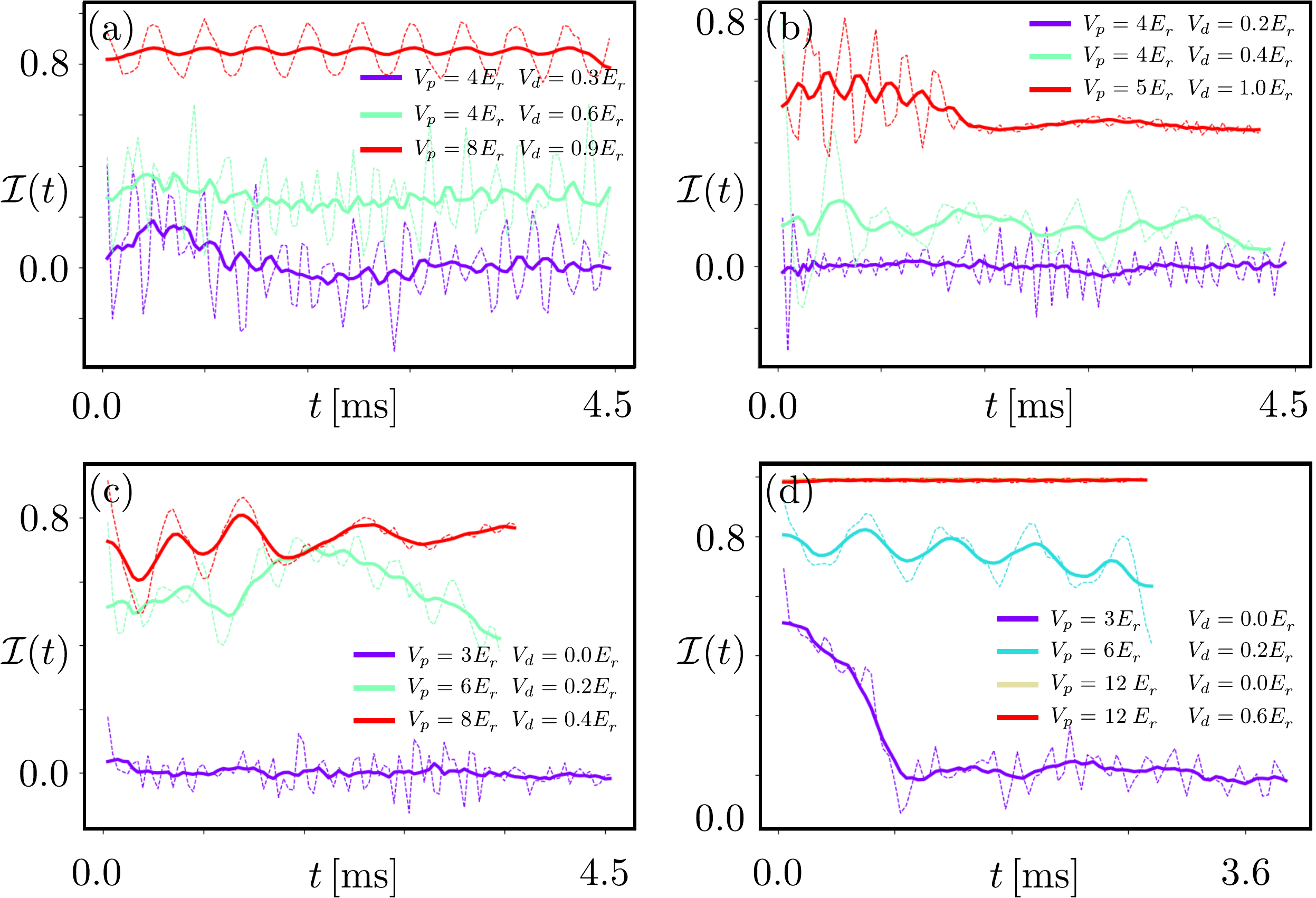}
\caption{
Imbalance dynamics for three different parameter values corresponding to the extended phase (purple curves), the intermediate phase (cyan/blue curves) and the localized phases (olive curve: DSBS state, red curve: standard localized phase).
The different panels depict the dynamics for different values of dipolar interactions: (a) $W=0.0 E_r$, (b) $W=0.02 E_r$, (c) $W=0.2 E_r$, (d) $W=2.0 E_r$.
The dashed lines are the raw data, while the thick solid lines are running averages.
}
\label{fig:imbalance}
\end{figure}

Fig.~\ref{fig:imbalance} shows the behavior of the density imbalance, which measures the wave function localization.
Each panel shows three or four representative parameter choices in the extended, intermediate, and localized phase, respectively, and for a different value of dipolar interaction strength: (a) $W=0.0 E_r$, (b) $W=0.02 E_r$, (c) $W=0.2 E_r$, (d) $W=2.0 E_r$.
The imbalance is an inherently noisier quantity because of the ambiguity of defining lattice sites in a continuum optical lattice.
To compensate for that, in selecting the value used to classify the different phases, we typically consider averages over the last 50 time steps.
The averages nicely recollect three distinct cases.
iI the extended phase, the initial imbalance rapidly decays towards zero and then oscillates close to that value for all subsequent times.
In the localized phase, the initial imbalance is mostly retained with a few oscillations. 
In particular, strong interactions preserve much better the initial imbalance [cf. Fig.~\ref{fig:imbalance}(d)].
In the intermediate phase, the imbalance exhibits a larger reduction from its initial value, but remains well above zero throughout the dynamics.

\newpage
{\color{white}.}
\newpage
\section{More details on the density expansion dynamics}
\begin{figure}[h!]
\centering
\includegraphics[width=1.0\columnwidth]{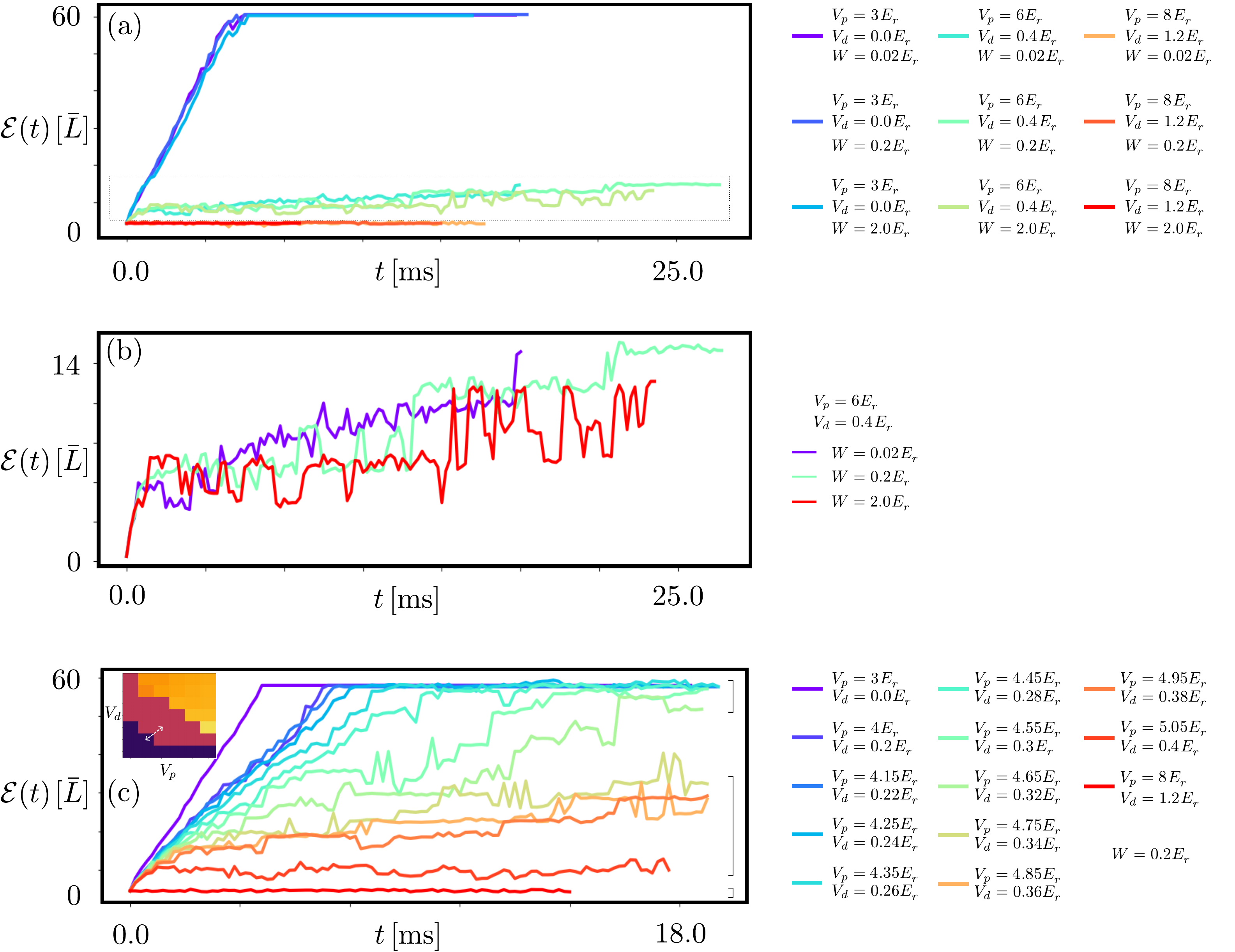}
\caption{
Expansion dynamics for in various regimes, measured as $\mathcal{E}(t)$.
(a): expansion dynamics of the nine points in parameter space analyzed in the main text. The dotted region is plotted in higher detail in panel (c).
(b): expansion dynamics in higher resolution across the extended-intermediate transition, shown with dashed white arrow in the phase diagram in the inset. For comparison, a data point in the deep extended region at $V_p=1.2 E_r^p$, $V_d = 0.0 E_r^p$ (purple curve) and deep in the localized region at $V_p=4.05 E_r^p$, $V_d = 0.61 E_r^p$ (dark red curve) are shown. The square brackets qualitatively separate the different phases based on their long-time limit expansion.
(c): Expansion dynamics of the intermediate phase with increasing dipolar interaction strength.
}
\label{fig:expansion-detailed}
\end{figure}

In this section, we present additional results for the density expansion dynamics that complement the indirect expansion measure obtained from the calculation of the edge density and illustrated in the main text.
In Fig. \ref{fig:expansion-detailed}, we show the behavior of the density expansion in time $\mathcal{E}(t)$ for various points in parameter space belonging to different phases.
The expansion is calculated as 
\begin{equation}
\mathcal{E}(t) \equiv \: \mathcal{M}_{>} [ \rho(x,t)] - \mathcal{M}_{<}[ \rho(x,t)]  - 2 \mathcal{M}_{>} [ \rho(x,0) ],
\end{equation}
where $\mathcal{M}_{>}$ ($\mathcal{M}_{<}$) calculates the point $x>0$ ($x<0$) where $\rho(x,t)$ has decayed below a certain threshold $\epsilon$, which we empirically define to be $\epsilon = 0.005$ in the plots below.
This indicates the additional spatial extent that the particles have reached at time $t$ when compared with the initial density.
Note that the initial density is completely symmetric with respect to the origin, therefore $\mathcal{M}_{<} [\rho(x,0)] = -\mathcal{M}_{>} [\rho(x,0)]$.
A value of $\mathcal{E}(t) \approx 0$ indicates low to no expansion.
A large value of $\mathcal{E}(t)$ indicates instead a large expansion.
Note that $\mathcal{E}(t)$ is bounded by the size of the simulation grid (64 $\bar{L}$).
Therefore, the expansion remains at that value if and when the density reaches the boundaries of the system.

From the figure \ref{fig:expansion-detailed}(a), we can distinguish three different behaviors for the data points analyzed in the main text and corresponding to the three different phases (extended, intermediate, and localized).
The extended phase shows a very rapid and linear expansion.
The localized phase, on the contrary, shows practically zero expansion.
The intermediate phase (highlighted by the grey rectangle) shows instead a slow, sublinear expansion that becomes progressively more chaotic as dipolar interactions increase.
Upon closer inspection in \ref{fig:expansion-detailed}(b), the intermediate phase shows a slight slowing down of expansion when interactions are increased.
This is a sign of the increasing interference of the tail of the dipole-dipole interactions.

Nevertheless, locating the phase boundary between extended and intermediate phase is not so easy based on the expansion dynamics alone. 
A pictorial demonstration of this is offered in \ref{fig:expansion-detailed}(c), where for $W=0.2 E_r$ we plot with higher resolution the expansion dynamics of the density for parameters across the transition from extended to localized phase (shown in the inset).
Indeed, while we progressively move towards the intermediate phase, the monotonic and rapid density expansion continuously slows down, and starts to acquire a non-monotonic character.
These results point towards a crossover from extended to intermediate phase, rather than a sharp transition. 
Indeed, the hallmark of the intermediate phase is rather a coexistence of both extended and localized states.
In the long-time limit, the extended states will always expand indefinitely.
However, localized states will remain localized in the initial configuration.
This is where the measurement of the imbalance comes into play to determine which regions of parameter space belong to the intermediate phase.

\newpage
{\color{white}.}
\newpage
\section{Behavior of average kinetic and interaction energy}
\begin{figure}[h!]
\centering
\includegraphics[width=\columnwidth]{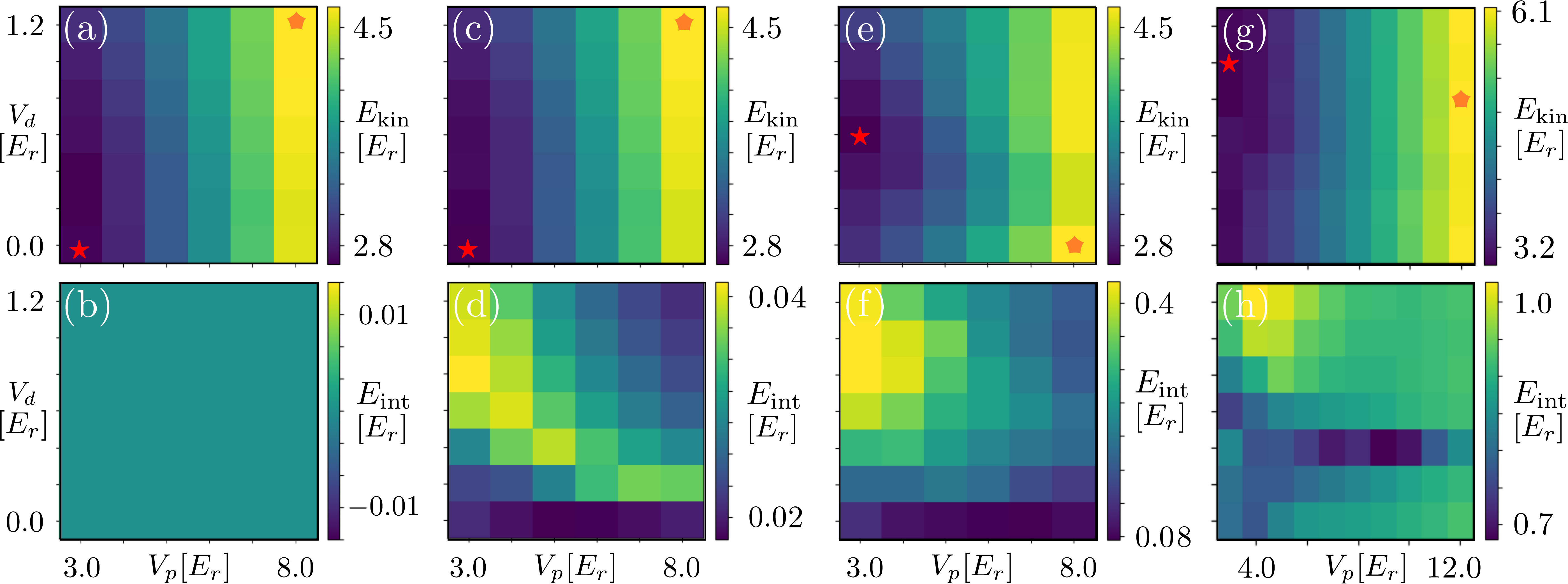}
\caption{
Average kinetic energy (top row) and interaction energy (bottom row) as a function of $V_p$ and $V_d$ for increasing dipolar interaction strength $W$.
(a)-(b): $W=0.0 E_r$.
(c)-(d): $W=0.02 E_r$.
(e)-(f): $W=0.2 E_r$.
(g)-(h): $W=2.0 E_r$.
The red stars and orange pentagons respectively indicate the minimum and maximum values of the kinetic energy in each diagram.
}
\label{fig:energies-p-3}
\end{figure}

In this section we examine the energetics of the interacting fermions in the quasiperiodic potential.
Fig.~\ref{fig:energies-p-3} depicts the behavior of the kinetic energy (top panels) and of the interaction energy (bottom panels) in the probed parameter space for increasing dipolar interaction strength $W$.
From the figures, we can see that the primary lattice depth $V_p$ is what drives most of the increase in the kinetic energy, but a region roughly delineating the intermediate phase appears as a faint ``shadow'' of points with slightly lower kinetic energy than their neighbors [see in particular Fig.~\ref{fig:energies-p-3}(e)].
Note also that the interactions modify the location of the maximal and minimal kinetic energy in parameter space.
Whereas the maximum (minimum) in the noninteracting system is deep in the localized (extended) phase, interactions push them towards other regions in a nonmonotonic fashion.

Contrary to expectations, the interaction energy does not seem to be strongly connected with the shapes appearing in the phase diagram. 
Nevertheless, the regions with strongest interaction energy systematically fall within the intermediate phase and are progressively pushed to deeper detuning lattice depths as $W$ is increased [Fig.~\ref{fig:energies-p-3}(d),(f),(h)]. 
This seems to indicate that strong quasicrystalline structures are the preferential setting to amplify the effect of long-range interactions.

It is also interesting to study the appearance of the regions with the \emph{lowest} interaction energy, indicating parameter ranges where the (quasi)crystalline structure of the potential competes strongly with the long-range interactions.
For weak to moderate dipolar interactions, the long-range repulsions are mostly inhibited in the extended phase, and in particular in the clean case (i.e. $V_d=0.0$).
This might be due to the fact that the repulsions tend to stabilize a periodic structure (a crystal state) that is different than the underlying optical lattice periodicity.
If the dipolar interactions are not strong enough ($W \le 0.2 E_r$), the extended eigenstates of the (quasi)periodic lattice dominate.
However, when they become stronger, the highly localized crystal state is energetic enough to compete and win over the extended eigenstates of the (quasi)periodic lattice.
This lifts the minimum of the interaction energy from $V_d=0.0 E_r$ to $V_d=0.4 E_r$.
This region precisely coincides with the location of the long, resonance-like lobe of the intermediate phase, where long-lived density oscillations occur at the boundaries.
It thus appears that the quasiperiodic structure and the dipolar interactions enter a kind of interference or resonance phenomenon at $V_d = 0.4 E_r$ that increases the stability of some of the extended states (which undergo pinned oscillations) even more than the clean lattice case.

\newpage
\section{Correlation dynamics}
\begin{figure}[h!]
\centering
\includegraphics[width=0.7\columnwidth]{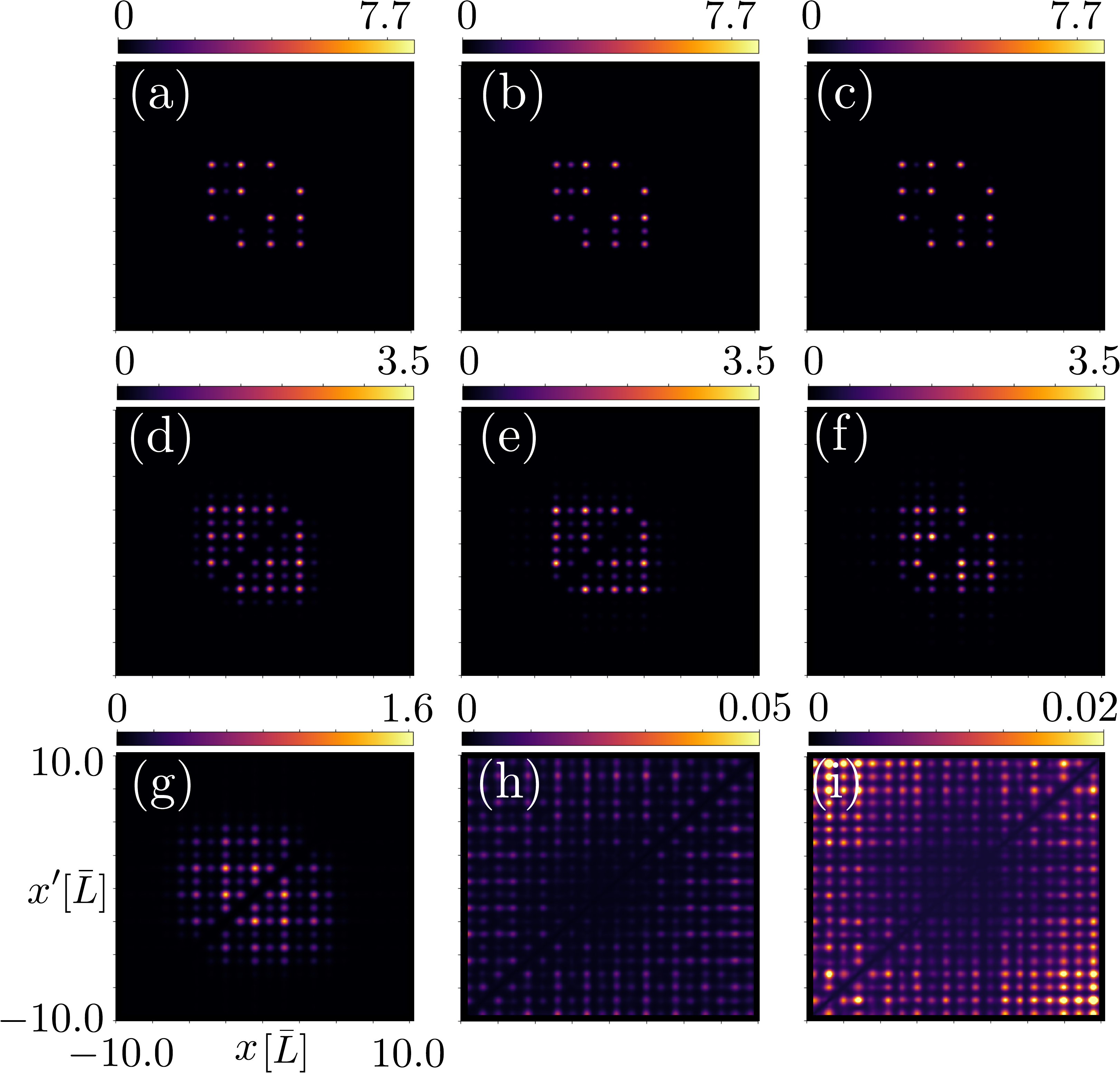}
\caption{
2-RDM for $N=4$ noninteracting fermions in the localized phase ($V_p=8.0 E_r$, $V_d=1.2 E_r$, top row), intermediate phase ($V_p=6.0 E_r$, $V_d=0.4 E_r$, middle row) and extended phase ($V_p=3.0 E_r$, $V_d=0 E_r$, bottom row).
For each phase, the 2-RDM is plotted at three different times.
The times are $t=0.089$ ms ((a), (d), (g)), $t=0.89$ ms ((b), (e), (h)), $t=2.23$ ms ((c), (f), (i)).
}
\label{fig:2-RDM-noninteracting-full}
\end{figure}

\begin{figure}[h!]
\centering
\includegraphics[width=0.7\columnwidth]{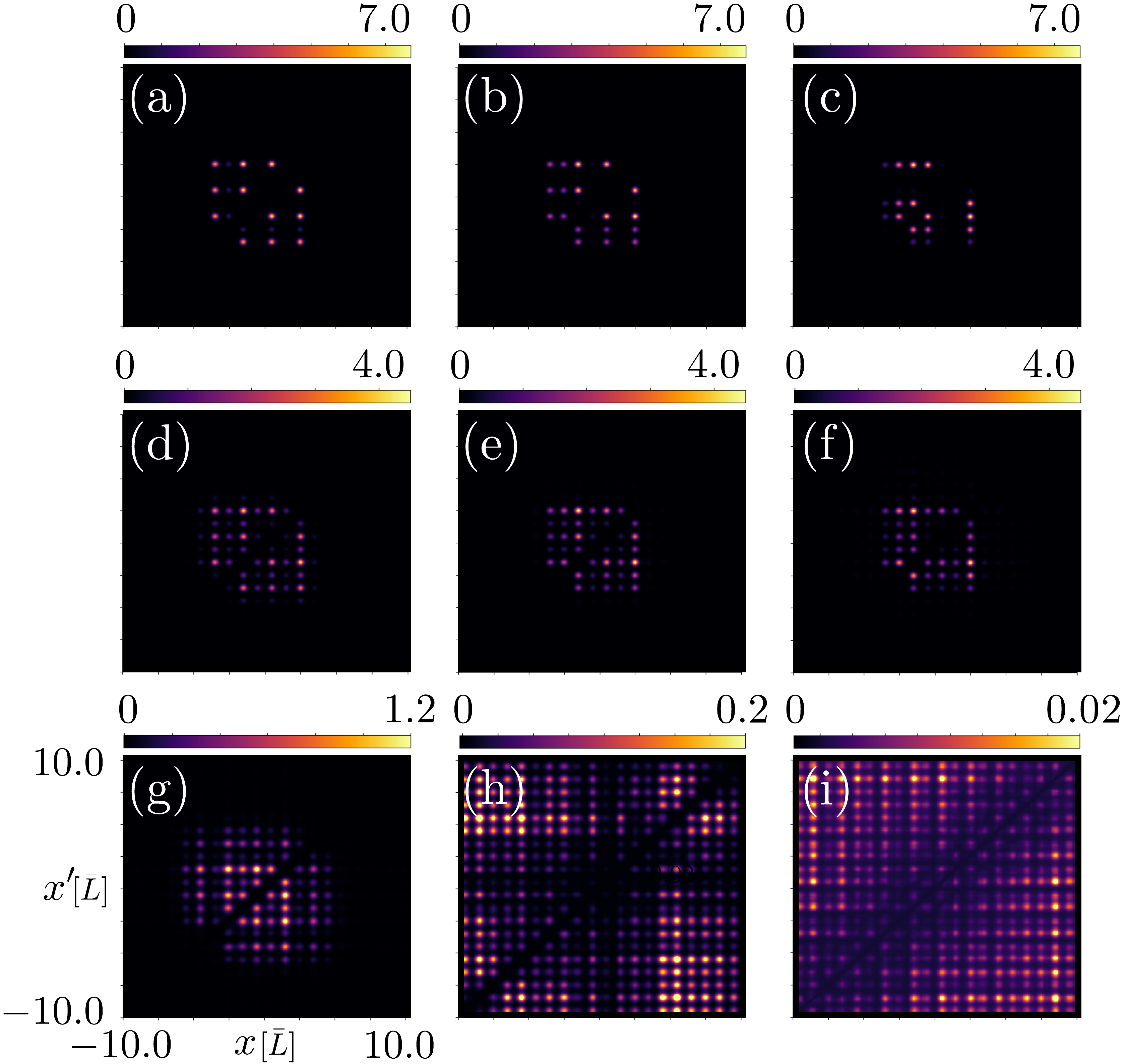}
\caption{
2-RDM for $N=4$ dipolar interacting fermions with $W=0.02 E_r$ in the localized phase ($V_p=8.0 E_r$, $V_d=1.2 E_r$, top row), intermediate phase ($V_p=6.0 E_r$, $V_d=0.6 E_r$, middle row) and extended phase ($V_p=4.0 E_r$, $V_d=0 E_r$, bottom row).
For each phase, the 2-RDM is plotted at three different times.
The times are $t=0.089$ ms ((a), (d), (g)), $t=0.89$ ms ((b), (e), (h)), $t=2.23$ ms ((c), (f), (i)).
}
\label{fig:corr-xl-0.1}
\end{figure}

\begin{figure}[h!]
\centering
\includegraphics[width=0.7\columnwidth]{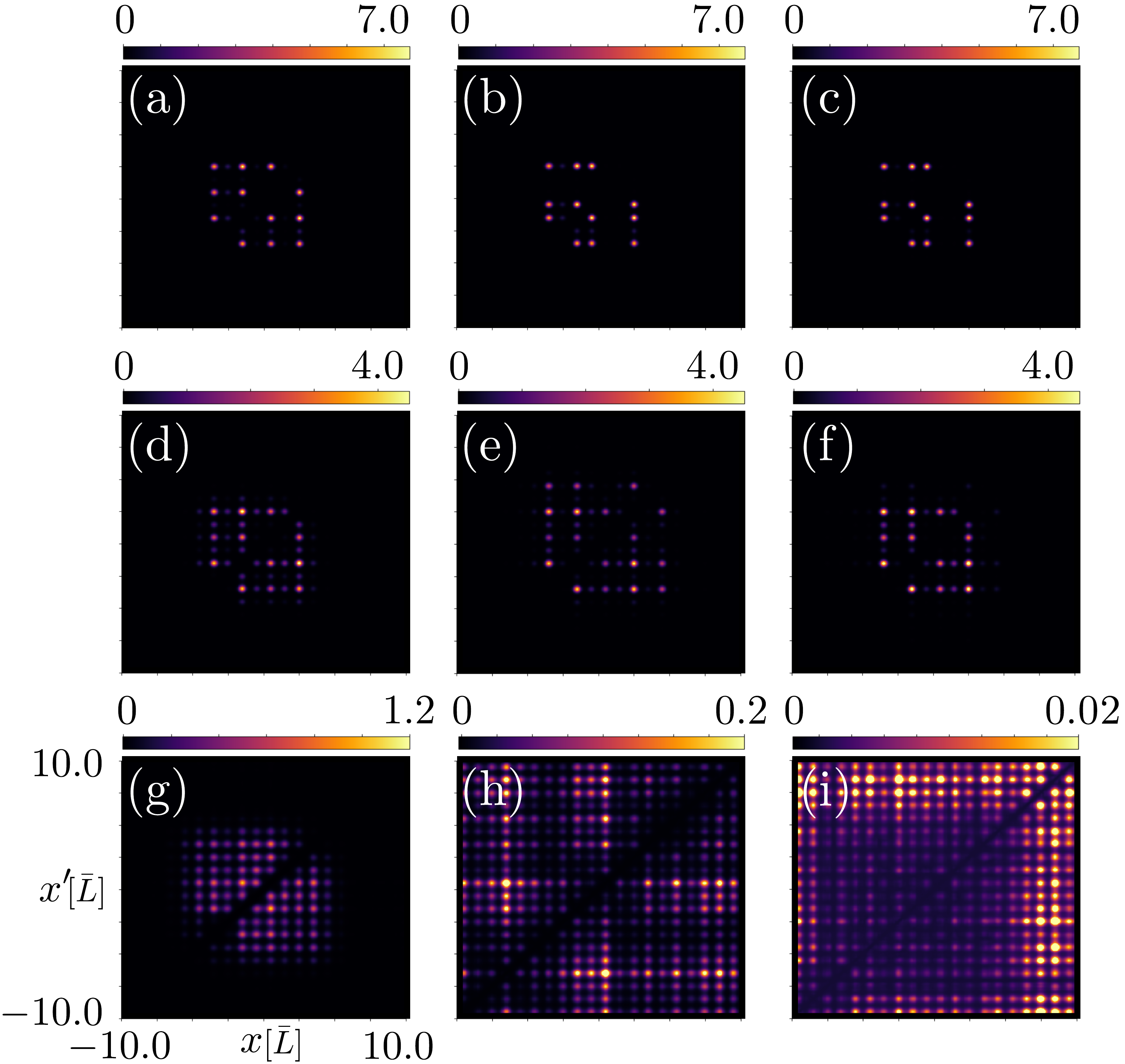}
\caption{
2-RDM for $N=4$ dipolar interacting fermions with $W=0.2 E_r$ in the localized phase ($V_p=8.0 E_r$, $V_d=1.2 E_r$, top row), intermediate phase ($V_p=6.0 E_r$, $V_d=0.6 E_r$, middle row) and extended phase ($V_p=4.0 E_r$, $V_d=0 E_r$, bottom row).
For each phase, the 2-RDM is plotted at three different times.
The times are $t=0.089$ ms ((a), (d), (g)), $t=0.89$ ms ((b), (e), (h)), $t=2.23$ ms ((c), (f), (i)).
}
\label{fig:corr-xl-1}
\end{figure}

\begin{figure}[h!]
\centering
\includegraphics[width=0.7\columnwidth]{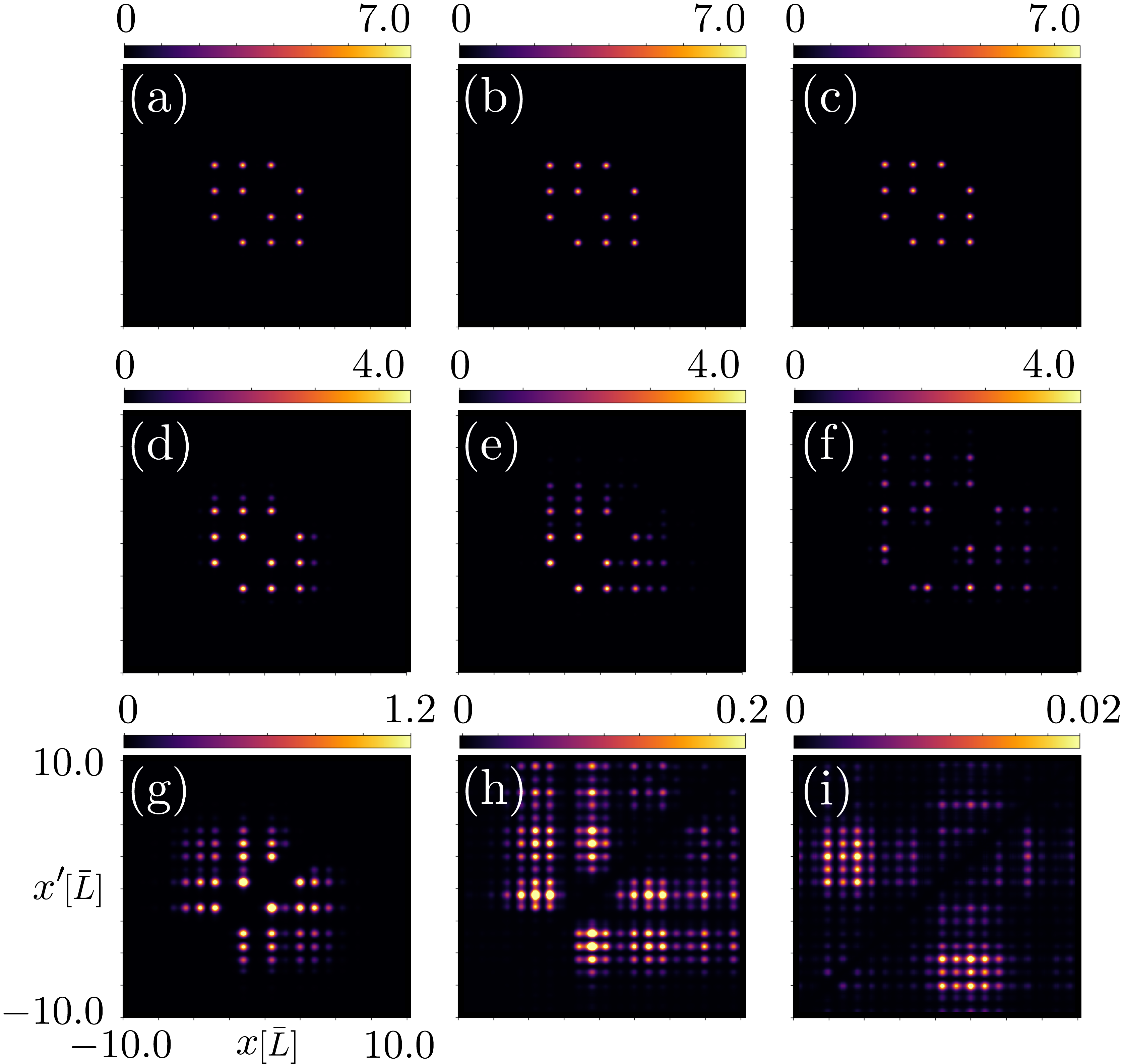}
\caption{
2-RDM for $N=4$ dipolar interacting fermions with $W=2.0 E_r$ in the localized phase ($V_p=8.0 E_r$, $V_d=1.2 E_r$, top row), intermediate phase ($V_p=6.0 E_r$, $V_d=0.6 E_r$, middle row) and extended phase ($V_p=4.0 E_r$, $V_d=0 E_r$, bottom row).
For each phase, the 2-RDM is plotted at three different times.
The times are $t=0.089$ ms ((a), (d), (g)), $t=0.89$ ms ((b), (e), (h)), $t=2.23$ ms ((c), (f), (i)).
}
\label{fig:corr-xl-10}
\end{figure}

In this section we present additional information about the dynamics of correlations in each phase and for increasing dipolar interaction strengths.
These figures complement the ones in the main text by providing information at earlier and later times ($t=0.089$ ms and $t=2.23$ ms).
The 2-RDM for the noninteracting case ($W=0$) and for three values of the interactions ($W=0.01$$E_r^p$, $W=0.10$$E_r^p$, and $W=1.01$$E_r^p$) is depicted in Figs.~\ref{fig:2-RDM-noninteracting-full}, \ref{fig:corr-xl-0.1}, \ref{fig:corr-xl-1}, and \ref{fig:corr-xl-10}, respectively.
All calculations were performed with $M=9$ orbitals.

The additional panels at earlier and later times reinforce the picture emerging from the figures in the main text.
Fermions in the localized phase exhibits little to no correlations with sites not occupied in the initial configurations.
In the extended phase their correlations rapidly spread to all sites with a rapid expansion.
In the intermediate phase, instead, the correlations exhibit a hybrid behavior between the other two cases: a core of superlattice correlations persists at longer times, but correlations slowly build up with other sites in between and expand to outer sites.
Furthermore, from the additional plots at longer times it becomes clear that increasing the strength of the dipolar repulsions slows down the dynamics and the expansion of the particles in the intermediate and extended phases by stabilizing localization.

\end{document}